\pdfoutput=1

\RequirePackage{fix-cm}

\documentclass[twocolumn]{svjour3}

\smartqed

\usepackage{graphicx}
\usepackage{amssymb,amsmath}
\usepackage{txfonts}
\usepackage[numbers,sort&compress]{natbib}
\usepackage{hyperref}

\journalname{Nonlinear Dynamics}

\begin{document}

\title{Orbit classification in the Hill problem - I. The classical case}

\author{Euaggelos E. Zotos}

\institute{Department of Physics, School of Science, \\
Aristotle University of Thessaloniki, \\
GR-541 24, Thessaloniki, Greece\\
Corresponding author's email: {evzotos@physics.auth.gr}}

\date{Received: 8 November 2016 / Accepted: 17 March 2017 / Published online: 27 March 2017}

\titlerunning{Orbit classification in the Hill problem - I. The classical case}

\authorrunning{Euaggelos E. Zotos}

\maketitle

\begin{abstract}

The case of the classical Hill problem is numerically investigated by performing a thorough and systematic classification of the initial conditions of the orbits. More precisely, the initial conditions of the orbits are classified into four categories: (i) non-escaping regular orbits; (ii) trapped chaotic orbits; (iii) escaping orbits; and (iv) collision orbits. In order to obtain a more general and complete view of the orbital structure of the dynamical system our exploration takes place in both planar (2D) and the spatial (3D) version of the Hill problem. For the 2D system we numerically integrate large sets of initial conditions in several types of planes, while for the system with three degrees of freedom, three-dimensional distributions of initial conditions of orbits are examined. For distinguishing between ordered and chaotic bounded motion the Smaller ALingment Index (SALI) method is used. We managed to locate the several bounded basins, as well as the basins of escape and collision and also to relate them with the corresponding escape and collision time of the orbits. Our numerical calculations indicate that the overall orbital dynamics of the Hamiltonian system is a complicated but highly interested problem. We hope our contribution to be useful for a further understanding of the orbital properties of the classical Hill problem.


\end{abstract}

\section{Introduction}
\label{intro}

The process where test particles escape from Hamiltonian systems is surely one of the most intriguing problems in nonlinear dynamics (e.g., \cite{C90,CK92,CKK93,SHA03,STN02}). In the case where the energy of escape has a finite value escape is possible. In particular, when the particles have values of energy higher than the critical energy of escape they are able to find the openings, or exit channels, in the equipotential surface and therefore escape to infinity. Systems with exit channels are also known as open or leaking Hamiltonian systems. The literature is replete with studies on such open Hamiltonian systems (e.g., \cite{BBS09,CHLG12,EP14,KSCD99,LT11,NH01,SCK95,SKCD95,SKCD96,Z14a,Z14b,Z15b,Z16b}).

The topic of leaking Hamiltonian systems however is by far less explored than the closely related issue of chaotic scattering. In many previous works the chaos theory has been successfully used in order to investigate and explain the phenomenon of chaotic scattering (e.g., \cite{BOG89,BTS96,BST98,DJT12,DGJT14,DJ16,GDJ14,J87,JS87,JP89,JR90,JT91,JLS99,JMS95,JMSZ10,SASL06,SSL07,SS08,SHSL09,SS10}). Needles to say that all the above-mentioned citations on both issues of chaotic scattering and open Hamiltonian systems are exemplary rather than exhaustive.

The restricted three-body problem (RTBP for short) is an excellent example of an open Hamiltonian system with escapes (e.g., \cite{WM94a,WM94b}). Over the last decade or so a large number of studies have been devoted on orbit classification in the RTBP. It all started with the pioneer works of Nagler \cite{N04,N05} where initial conditions of orbits were classified as bounded, escaping or collision. Moreover, bounded orbits were further classified into orbital families by taking into account the type of the motion of the test particle around the primary bodies. In the same vein, orbit classification in the RTBP with perturbations has also been performed in \cite{Z15c,Z15d} where we investigated the influence of the oblateness, while in \cite{Z15e} we explored how the radiation pressure affects the orbital content of the dynamical system.

Similarly, orbit classification has also been conducted in celestial mechanics and dynamical astronomy regarding planetary systems. In particular, the orbital dynamics of the Earth-Moon system with two degrees of freedom using a scattering region around the Moon has been studied in \cite{dAT14}. In \cite{Z16a} the numerical investigation has been expanded into three dimensions thus exploring the orbital structure of the three degrees of freedom Earth-Moon system. Furthermore, very recently the escape and collision dynamics in the Saturn-Titan and in the Pluto-Charon binary planetary systems have been revealed in \cite{Z15f} and \cite{Z15g}, respectively, by classifying initial conditions of orbits in the configuration space.

The Hill limiting case is in fact a simplified modification of the RTBP which focus on the vicinity of the secondary (e.g., \cite{H86,PH86,PH87}). This allow us to study the motion of the test particles in the neighborhood of the equilibrium points $L_1$ and $L_2$. At this point it should be emphasized that the Hill approximation is valid only when the mass of the secondary is much smaller compared with the mass of the primary body. One can directly obtain the Hill model from the classical RTBP by translating the origin to the center of the secondary and also by rescaling the coordinates by a factor $\mu^{1/3}$. In \cite{SS00} the authors performed an extensive investigation of the planar Hill problem by computing several homoclinic and heteroclinic connections. In this work, which is the first part of a series of papers, we shall perform a thorough and systematic orbit classification in the classical Hill problem. To our knowledge, there are no previous detailed and systematic numerical studies regarding orbit classification in the classical Hill problem. On this basis, the new information presented in this paper is exactly the contribution as well as the novelty of this research work. In the following papers of the series we will explore how several perturbations (i.e., the oblateness and the radiation pressure) influence the orbital structure of the dynamical system.

The present paper is organized as follows: In Section \ref{mod} we present in detail the derivation and the basic dynamical properties of the mathematical model. All the computational techniques we used in order to classify the initial conditions of the orbits are described in Section \ref{cometh}. In the following Section, a thorough and systematic numerical investigation takes place which allow us to reveal the overall orbital structure of the Hill system with two and three degrees of freedom. Our paper ends with Section \ref{disc} where the discussion and the main conclusions of our research are given.

\section{Derivation of the mathematical model}
\label{mod}

The classical Hill problem is derived from the circular restricted three-body problem if we make some scale changes and if we also take the limiting case where the mass ratio $\mu = m_2/(m_1 + m_2)$ tends to zero $(\mu \rightarrow 0)$.

The time-independent effective potential function of the circular restricted three-body problem is
\begin{equation}
\Omega(X,Y,Z) = \frac{\left(1 - \mu \right)}{r_1} + \frac{\mu}{r_2} + \frac{1}{2}\left(X^2 + Y^2 \right),
\label{pot}
\end{equation}
where $(X,Y,Z)$ are the coordinates of the test particle, while
\begin{align}
r_1 &= \sqrt{\left(X - \mu\right)^2 + Y^2 + Z^2}, \nonumber\\
r_2 &= \sqrt{\left(X - \mu + 1\right)^2 + Y^2 + Z^2},
\end{align}
are the distances of the test particle from the two primaries.

According to \cite{S67}, in a rotating system of reference the equations of motion are
\begin{align}
\Omega_X &= \frac{\partial \Omega}{\partial X} = \ddot{X} - 2\dot{Y}, \nonumber\\
\Omega_Y &= \frac{\partial \Omega}{\partial Y} = \ddot{Y} + 2\dot{X}, \nonumber\\
\Omega_Z &= \frac{\partial \Omega}{\partial Z} = \ddot{Z}.
\end{align}

We now place the origin of the coordinates at the center of the secondary and we change the scale of lengths by a factor of $\mu^{1/3}$
\begin{equation}
X = \mu - 1 + \mu^{1/3}x, \ \ \ Y = \mu^{1/3}y, \ \ \ Z = \mu^{1/3}z.
\label{trans0}
\end{equation}
Here we would like to note that the notations $(X,Y,Z)$ and $(x,y,z)$, regarding the coordinates for the RTBP and the Hill problem, respectively are according to \cite{PMD05}.

Applying the above-mentioned transformations to (\ref{pot}) we obtain
\begin{equation}
\frac{1}{\mu^{2/3}}\left(\Omega - \frac{3}{2}\right) = \frac{3x^2}{2} - \frac{z^2}{2} + \frac{1}{r} + \mathcal{O}(\mu^{1/3}),
\label{trans}
\end{equation}
with $r = \sqrt{x^2 + y^2 + z^2}$.

Taking the limit of the right-hand side of (\ref{trans}) for $\mu \rightarrow 0$ we arrive at the potential function of the classical Hill problem
\begin{equation}
W(x,y,z) = \frac{3x^2}{2} - \frac{z^2}{2} + \frac{1}{r}.
\label{pot2}
\end{equation}

More details about the derivation of the potential function of the Hill problem are given in the Appendix.

The corresponding equations of motion are
\begin{align}
W_x &= \frac{\partial W}{\partial x} = \left(3 - \frac{1}{r^3}\right)x = \ddot{x} - 2\dot{y}, \nonumber\\
W_y &= \frac{\partial W}{\partial y} = - \frac{y}{r^3} = \ddot{y} + 2\dot{x}, \nonumber\\
W_z &= \frac{\partial W}{\partial z} = - \left(1 + \frac{1}{r^3}\right)z = \ddot{z}.
\label{eqmot}
\end{align}

The Hill problem admits a Jacobi integral of motion
\begin{equation}
J = 2W(x,y,z) - \left(\dot{x}^2 + \dot{y}^2 + \dot{z}^2 \right),
\label{ham}
\end{equation}
where $J$ is the new Jacobi constant which is related to the Jacobi constant $C$ of the restricted three-body problem  through the relation
\begin{equation}
C = 3 + \mu^{2/3}J.
\end{equation}

\begin{figure*}[!t]
\centering
\resizebox{\hsize}{!}{\includegraphics{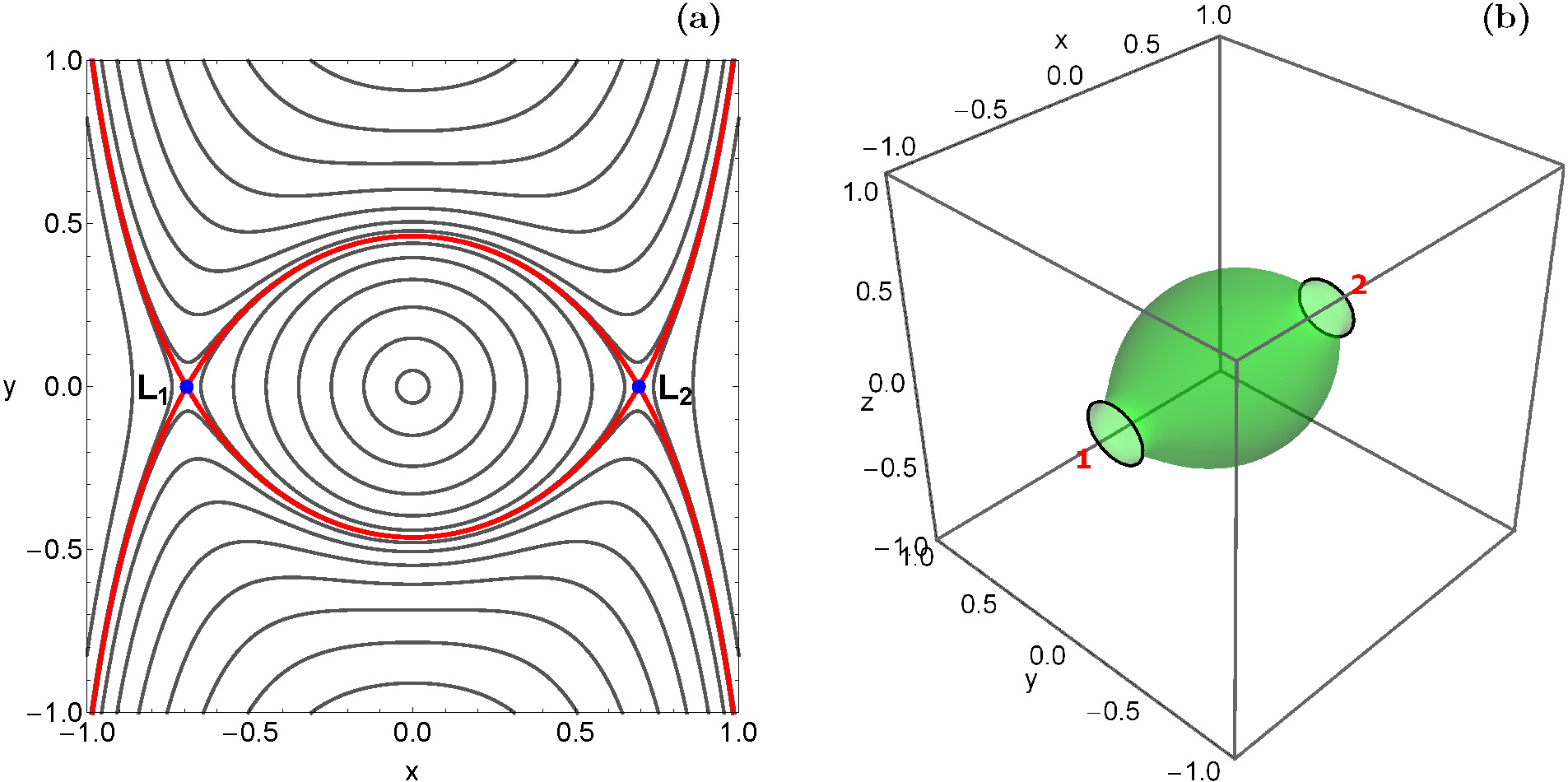}}
\caption{(a-left): The isoline contours of the effective potential function $W$ in the configuration $(x,y)$ plane $(z = 0)$. The positions of the two equilibrium points are indicated by blue dots. The isoline contours corresponding to the critical Jacobi value $J_L$ are shown in red. (b-right): The three-dimensional equipotential surface of the effective potential function $2W(x,y,z) = J$, when $J = 4.25$. The orbits can leak out through the exit channels 1 and 2 of the equipotential surface passing either through $L_1$ or $L_2$, respectively. (For the interpretation of references to colour in this figure caption and the corresponding text, the reader is referred to the electronic version of the article.)}
\label{conts}
\end{figure*}

In order to locate the positions of the equilibrium points of the system we have to solve the system of differential equations
\begin{equation}
W_x = W_y = W_z = 0.
\end{equation}
The Hill problem has only two equilibrium points which are located on the $x$ axis at $x_L = \pm 1/3^{1/3}$ (see Fig. \ref{conts}a). The value of the Jacobi integral at the equilibrium points is $J_L =  3^{4/3}$ and this is a critical value. This is because for $J < J_L$ the zero velocity surfaces are open and two symmetrical channels (exits) are present near the equilibrium points $L_1$ and $L_2$ through which the test particles can escape from the interior region $(-x_L \leq x \leq x_L)$ of the system. Therefore we may argue that $J_L$ plays, in a way, the role of the energy of escape and this is exactly why it is a critical value of the Jacobi integral. In Fig. \ref{conts}b we present a plot of the three dimensional equipotential surface $2W(x,y,z) = J$, when $J = 4.25$. We observe the presence of the two channels (or throats) through which the test particles can leak out. One could say that these two exit channels act, in a way, as hoses thus allowing the test particles to travel from the interior region of the system to the ``outside world". Channel 1 indicates escape towards the negative $x$ direction, while channel 2 indicates escape towards the positive $x$ direction.

\section{Computational methods}
\label{cometh}

In order to reveal the orbital structure of the Hill problem we need to define sets of initial conditions whose properties will be identified. For obtaining a more complete and spherical view regarding the nature of orbits we are going to explore both the two-dimensional (2D) and the three-dimensional (3D) Hill systems. In this work we decided to classify orbits with initial conditions in the configuration space only. In particular, for several values of the Jacobi integral $J$ we define dense uniform grids of initial conditions inside the corresponding zero velocity surface. For the 2D system $(z = 0)$, grids of $1024 \times 1024$ initial conditions $(x_0,y_0)$ are numerically integrated. For all 2D orbits $\dot{x_0} = 0$, while the initial value of $\dot{y}$ is always obtained from the Jacobi integral of motion (\ref{ham}), as $\dot{y_0} = \dot{y}(x_0,y_0,\dot{x_0},J) > 0$. In the same vein, for the 3D system, grids of $100 \times 100 \times 100$ initial conditions $(x_0,y_0,z_0)$ are numerically integrated. For all 3D orbits $\dot{x_0} = \dot{z_0} = 0$, while the initial value of $\dot{y}$ is always obtained from the energy integral (\ref{ham}), as $\dot{y_0} = \dot{y}(x_0,y_0,z_0,\dot{x_0},\dot{z_0},J) > 0$. In both cases, the initial conditions of the orbits lie inside the interior region (which is the scattering region) with $R = \sqrt{x_0^2 + y_0^2 + z_0^2} < x_L$.

The classification of the initial conditions of the orbits in the Hill problem is a rather demanding task if we take into account that the configuration space extends to infinity. In this study we shall classify initial conditions of orbits into four main categories:
\begin{enumerate}
  \item Orbits that move in bounded, or non-escaping regular orbits inside the interior region.
  \item Orbits that move in trapped chaotic orbits inside the interior region.
  \item Orbits that escape from the interior region passing either through $L_1$ or $L_2$.
  \item Orbits that collide with the secondary located at the origin of the coordinates.
\end{enumerate}
Our next task is to define appropriate numerical criteria in order to distinguish between the above-mentioned four types of motion. An orbit is considered to escape from the system when $x < -x_L - \delta$, or when $x > x_L + \delta$, where $\delta = 0.1$. At this point, it should be clarified that the tolerance $\delta$ was included so as to avoid the unstable Lyapunov orbits \cite{L49}, located near the equilibrium points, to be incorrectly classified as escaping orbits. A collision with the secondary occurs when $R < R_{\rm col}$, where $R_{\rm col} = 10^{-4}$.

Our preliminary calculations suggest that a considerable portion of the initial conditions correspond to bounded orbits, which can be either regular or chaotic. Therefore we decided to distinguish between initial conditions of regular non-escaping and trapped chaotic motion. Over the years many dynamical indicators have been introduced for distinguishing between regular and chaotic orbits. We chose to use the Smaller ALingment Index (SALI) method \cite{S01}, which has been proved a very fast and accurate tool. The mathematical definition of SALI is the following
\begin{equation}
\rm SALI(t) \equiv min(d_-, d_+),
\label{sali}
\end{equation}
where $d_- \equiv \| {\vec{w_1}}(t) - {\vec{w_2}}(t) \|$ and $d_+ \equiv \| {\vec{w_1}}(t) + {\vec{w_2}}(t) \|$ are the alignments indices, while ${\vec{w_1}}(t)$ and ${\vec{w_2}}(t)$, are two deviation vectors which initially point in two random directions. For distinguishing between ordered and chaotic motion, all we have to do is to compute the SALI along a time interval $t_{\rm max}$ of numerical integration. In particular, we track simultaneously the time-evolution of the main orbit itself as well as the two deviation vectors ${\vec{w_1}}(t)$ and ${\vec{w_2}}(t)$ in order to compute the SALI.

The determination of the nature of an orbit is obtained from the final value of the SALI at the end of the numerical integration. In particular, if SALI $> 10^{-4}$ the motion is regular, while if SALI $< 10^{-8}$ the motion is chaotic. If the final value of SALI lies in the interval $[10^{-4}, 10^{-8}]$ then we have the case of a ``sticky" orbit\footnote{Sticky orbits need an extremely long time interval in order to move away from the invariant sticky tori \cite{PW94}. Therefore, sticky orbits behave, for long time interval, as regular ones before revealing their true chaotic nature.} and further numerical integration is required for obtaining the true nature of the orbit. In our study the maximum time of the numerical integration $(t_{\rm max})$ was set to be equal to $10^4$ time units. Orbits that remain inside the interior region after integrating them for $t_{\rm max}$ are considered as non-escaping regular or trapped chaotic, according to their final value of SALI.

The set of the equations of motion (\ref{eqmot}) as well as the corresponding variational equations, needed for the computation of the SALI, have been forwarded integrated using a double precision Bulirsch-Stoer \verb!FORTRAN 77! algorithm (e.g., \cite{PTVF92}) with a variable time step. Throughout our numerical calculations the error regarding the conservation of the Jacobi integral of motion (\ref{ham}) was generally smaller than $10^{-12}$, while in some cases it was observed to be smaller than $10^{-14}$. For those initial conditions of orbits which move inside a region of radius $R < 10^{-2}$, thus leading to collision with the secondary, the Lemaitre's global regularization method (e.g., \cite{S67}) has been applied. All graphical illustrations presented in this paper have been created using the latest version 11 of the software Mathematica$^{\circledR}$ (e.g., \cite{W03}).

\section{Orbit classification}
\label{orbclas}

\begin{figure*}[!t]
\centering
\resizebox{\hsize}{!}{\includegraphics{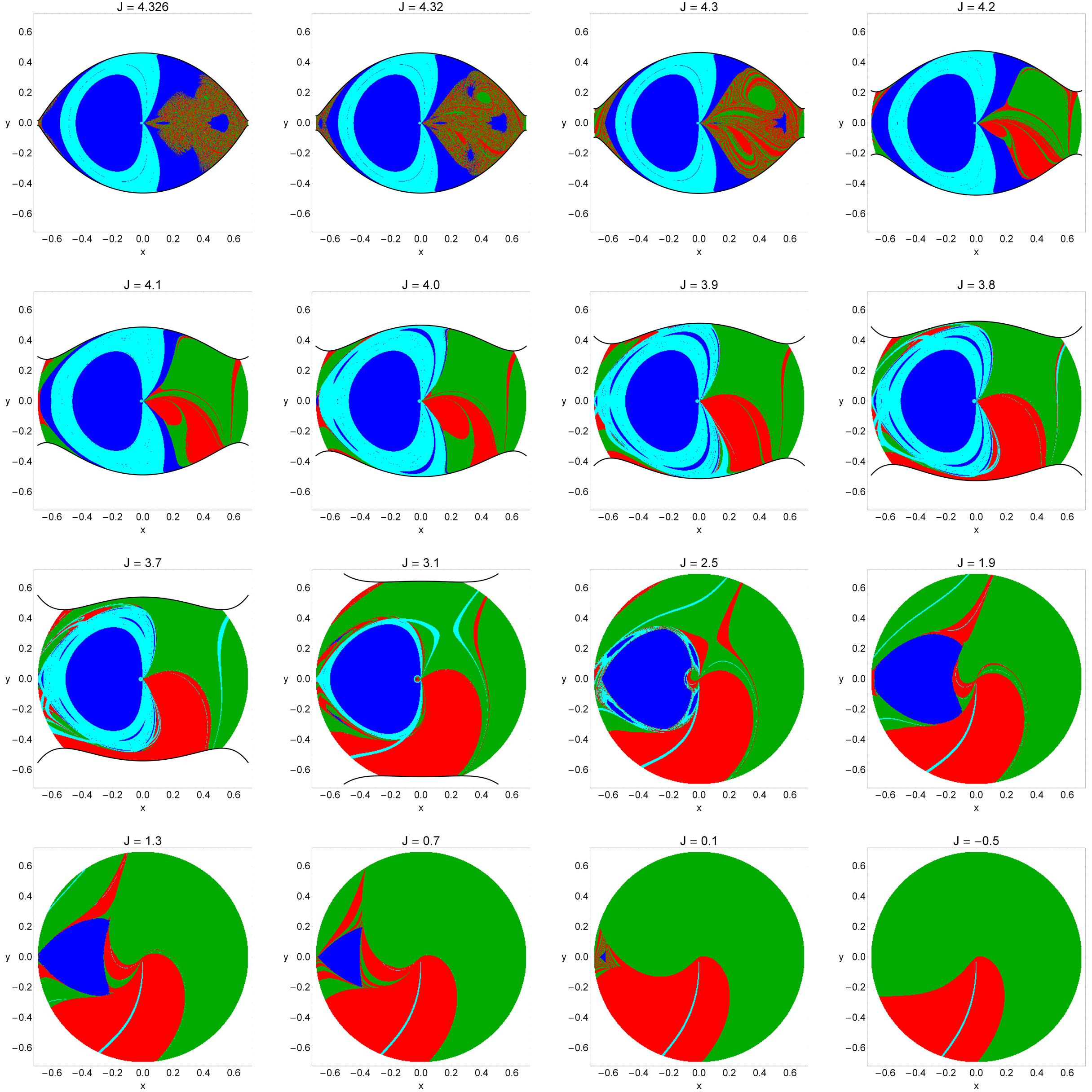}}
\caption{Evolution of the orbital structure of the configuration $(x,y)$ space with decreasing value of the Jacobi integral of motion. The color code is as follows: non-escaping regular orbits (blue), trapped chaotic orbits (yellow), escaping orbits through $L_1$ (red), escaping orbits through $L_2$ (green), and collision orbits (cyan). (For the interpretation of references to colour in this figure caption and the corresponding text, the reader is referred to the electronic version of the article.)}
\label{xy}
\end{figure*}

\begin{figure*}[!t]
\centering
\resizebox{\hsize}{!}{\includegraphics{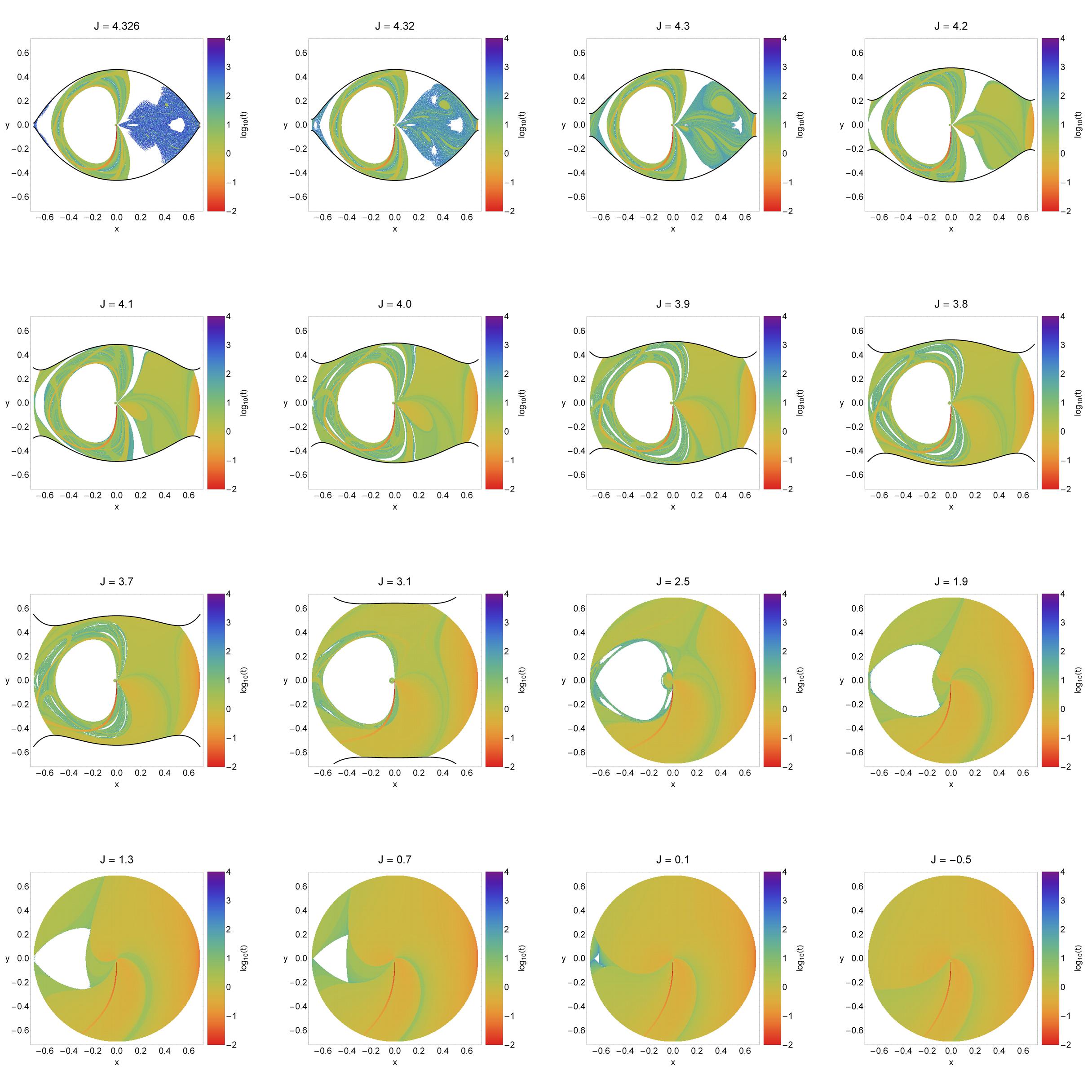}}
\caption{Distribution of the corresponding escape and the collision time of the orbits with initial conditions on the configuration $(x,y)$ space for the values of Jacobi integral of motion presented in Fig. \ref{xy}. The bluer the color, the larger the escape or the collision time. Initial conditions of both non-escaping regular orbits and trapped chaotic orbits are shown in white. (For the interpretation of references to colour in this figure caption and the corresponding text, the reader is referred to the electronic version of the article.)}
\label{xyt}
\end{figure*}

In this section we shall classify orbits with initial conditions in the configuration space into the four categories analyzed in the previous section. Parallel to the classification we shall also record the time scale (or time period) of the collision and the time scale of the escape. Our aim is to numerically explore the orbital properties of the Hill problem for several values of the Jacobi integral of motion.

In the following we shall present color-coded diagrams (CCDs) in which each pixel is assigned a specific color according to the particular type of the nature of the orbit. These CCDs are modern types of color-coded maps where the phase space is a complex mixture of basins of escape, collision basins and regions of bounded motion. By the term ``basin" we refer to a set of initial conditions which lead to a certain final state (collision, escape or bounded motion). Our preliminary numerical calculations indicate that bounded motion is almost always present. Generally speaking, the vast majority of the bounded basins correspond to initial conditions of regular orbits, where an adelphic integral of motion is present. This additional integral poses new restrictions to the available phase space and therefore it prevents them from escaping to infinity.

\subsection{Results for the 2D system}
\label{ss1}

We begin our investigation with the two-dimensional (2D) system when $z = \dot{z} = 0$. In Fig. \ref{xy} we present, using a rich collection of CCDs, the evolution of the orbital structure of the configuration $(x,y)$ plane as the value of the Jacobi constant decreases. The outermost black solid line denotes the zero velocity curve which is defined as $2W(x,y) = J$. We observe that when $J = 4.326$, that is an energy level just above the critical value $J_L$, the vast majority of the configuration plane is covered either by stability islands or collision basins. The remaining area is a highly fractal\footnote{When we state that an area is fractal we simply suggest that it has a fractal-like geometry without conducting any specific calculations regarding the fractal dimensions as in \cite{AVS09}.} mixture of escaping orbits. Inside this fractal area it is impossible to predict through which channel a test particle will escape. This is true because even a slight change on the initial conditions of an orbit leads the test particle to escape from the opposite channel, which is of course a classical indication of chaotic motion. As the value of the Jacobi integral of motion reduces even further (remember that at the same time the total orbital energy increases) the following important changes take place on the configuration $(x,y)$ plane:
\begin{itemize}
  \item The two symmetrical escape channels near the equilibrium points become more and more wide, as the energetically forbidden regions are reduced.
  \item Several basins of escape start to emerge inside the escape regions, while the fractality of the same regions is reduced. In particular, for $J < 4.2$ fractal regions are only present in the vicinity of the basin boundaries.
  \item The area of the basins of escape grows rapidly and for $J < 3.7$ it covers more than 50\% of the $(x,y)$ plane. It is observed that for relatively low values of the Jacobi integral of motion $(J < 0)$ escape basins dominate occupying almost the entire configuration space.
  \item The extent of the main stability island located at the left side of the $(x,y)$ plane $(x < 0)$ is gradually reduced and when $J < 0$ there is no indication of non-escaping regular motion whatsoever. Additional smaller stability islands appear and disappear with decreasing value of $J$.
  \item The collision basins are mainly located around the stability island. For $J < 2.5$ however, they appear only as thin filaments inside the vast escape domains. Our calculations suggest that collision motion remains possible even for extremely low values of the Jacobi value.
  \item With increasing orbital energy it is seen that the vast majority of the escaping orbits leak out through exit channel 2. In fact at the highest energy level studied $(J = -0.5)$ the area covered by initial conditions of orbits that escape through $L_2$ is almost four times larger than the area of initial conditions of orbits that escape through $L_1$. Therefore we may say that in the 2D system exit channel 2 seems to be much more preferable.
  \item Our computations indicate that in the 2D Hill system there is no numerical evidence of trapped chaotic motion.
\end{itemize}

In the following Fig. \ref{xyt} we demonstrate how the escape and the collision time of the orbits are distributed on the configuration $(x,y)$ space for the same values of the Jacobi integral shown in Fig. \ref{xy}. Light reddish colors correspond to fast escape or collision orbits, dark blue/purple colors indicate large escape and collision time, while white color denote regions of non-escaping regular motion and trapped chaotic motion. We note that the scale on the color bar is logarithmic varying always from -2 to 4. Inspecting the distribution of the escape time of orbits it is rather easy to associate the stable manifold of the non-attracting invariant set with medium escape time. Similarly, the largest observed escape rates are directly linked with the presence of sticky motion in the vicinity of boundaries of the several stability islands (see also \cite{CH10}). It is interesting to emphasize that when $J = 4.326$ most of the orbits escape only after 1000 time units. However as we proceed to higher energy levels the average escape time is considerably reduced. As it was seen in Fig. \ref{xy}, for $J < 2.5$ only a thin filament of collision orbits is presents. Our calculations suggest that these orbits collide within the first ten steps of the numerical integration.

\begin{figure}[!t]
\centering
\resizebox{\hsize}{!}{\includegraphics{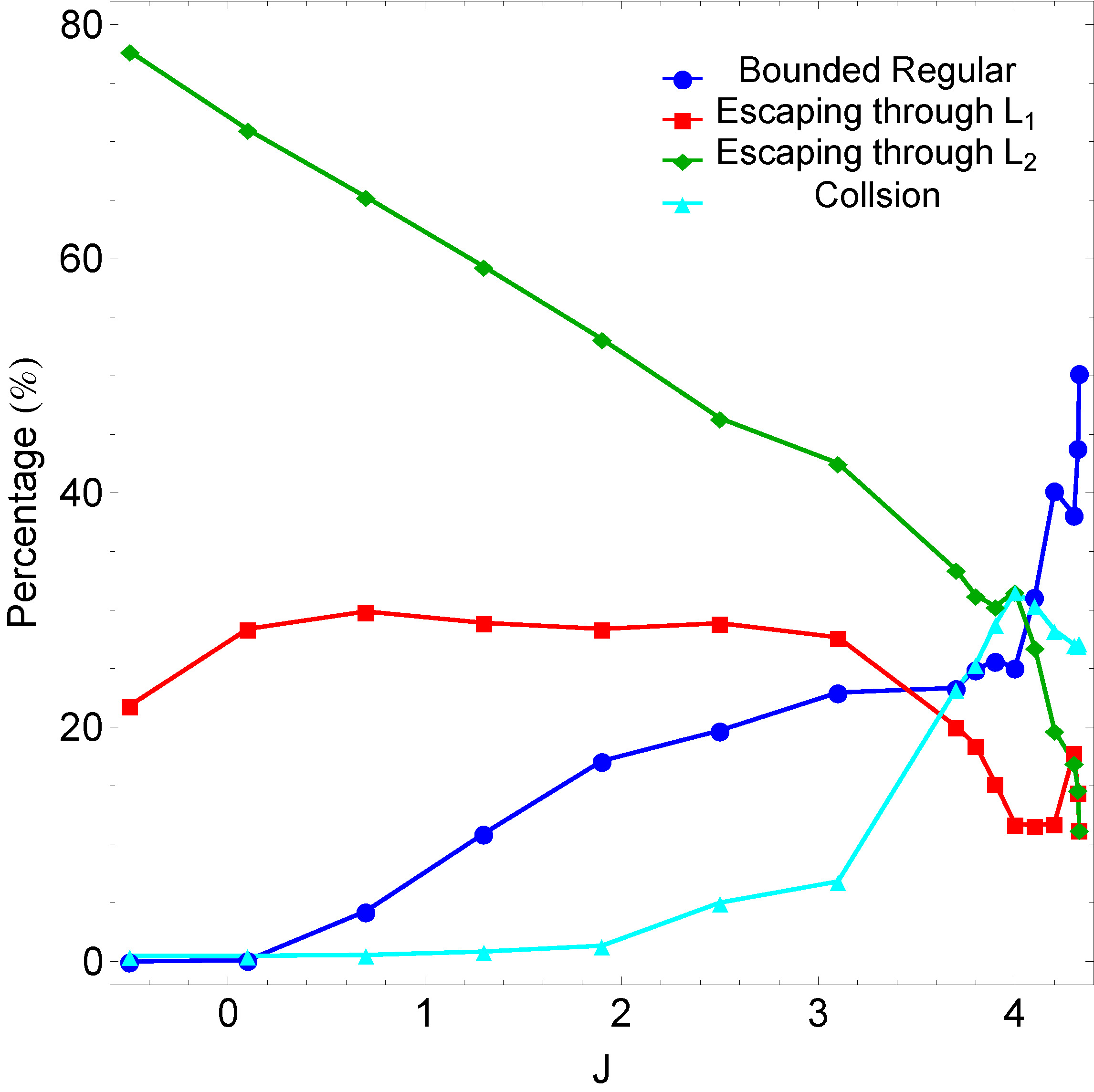}}
\caption{Evolution of the percentages of all types of orbits with initial conditions in the configuration $(x,y)$ plane, as a function of the Jacobi constant $J$. (For the interpretation of references to colour in this figure caption and the corresponding text, the reader is referred to the electronic version of the article.)}
\label{pxy}
\end{figure}

\begin{figure*}[!t]
\centering
\resizebox{\hsize}{!}{\includegraphics{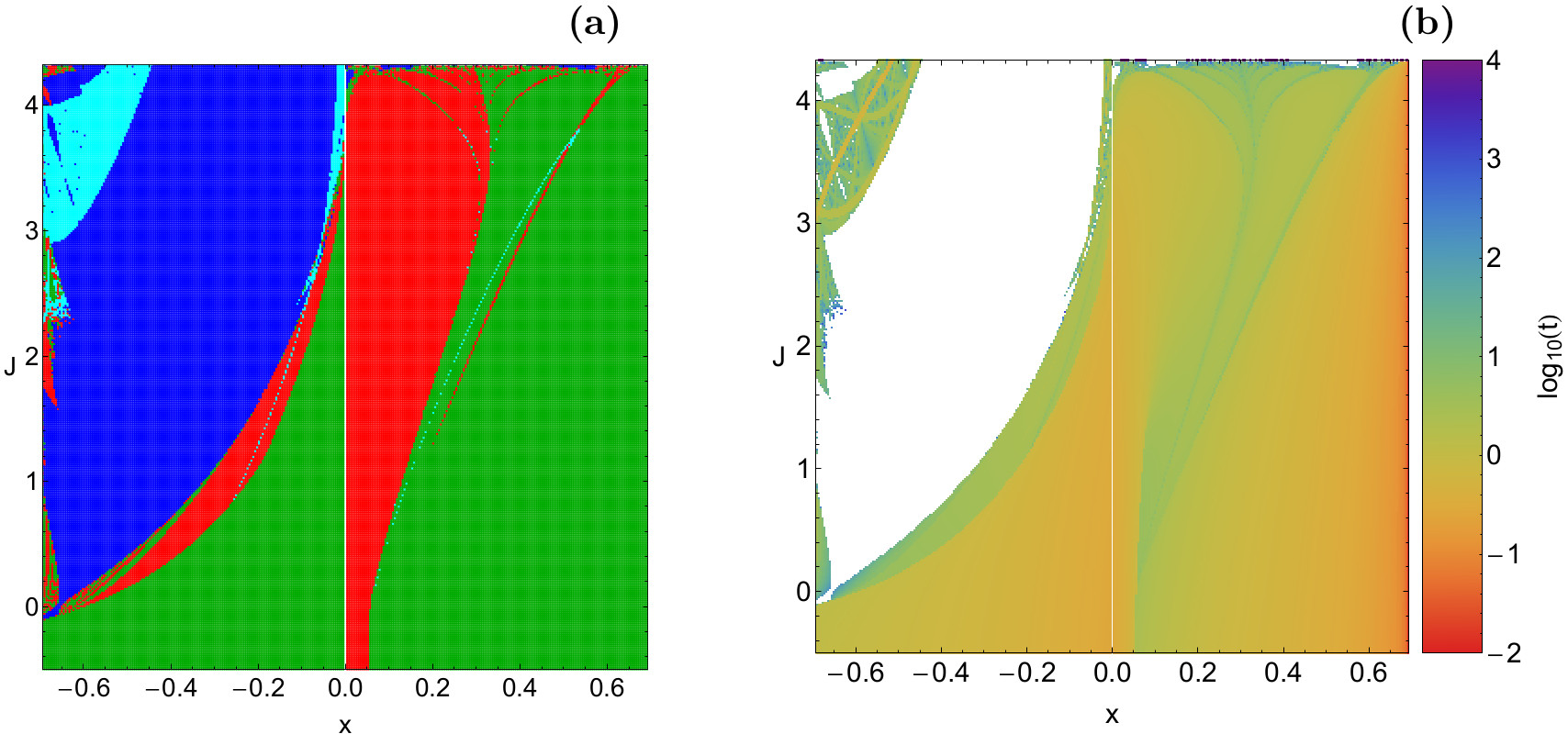}}
\caption{(a-left): Orbital structure of the $(x,J)$ plane when $J \in [-0.5,J_L)$. The color code is the same as in Fig. \ref{xy}. (b-right): Distribution of the corresponding escape and collision time of the orbits with initial conditions on the $(x,J)$ plane. (For the interpretation of references to colour in this figure caption and the corresponding text, the reader is referred to the electronic version of the article.)}
\label{xCt}
\end{figure*}

\begin{figure}[!t]
\centering
\resizebox{\hsize}{!}{\includegraphics{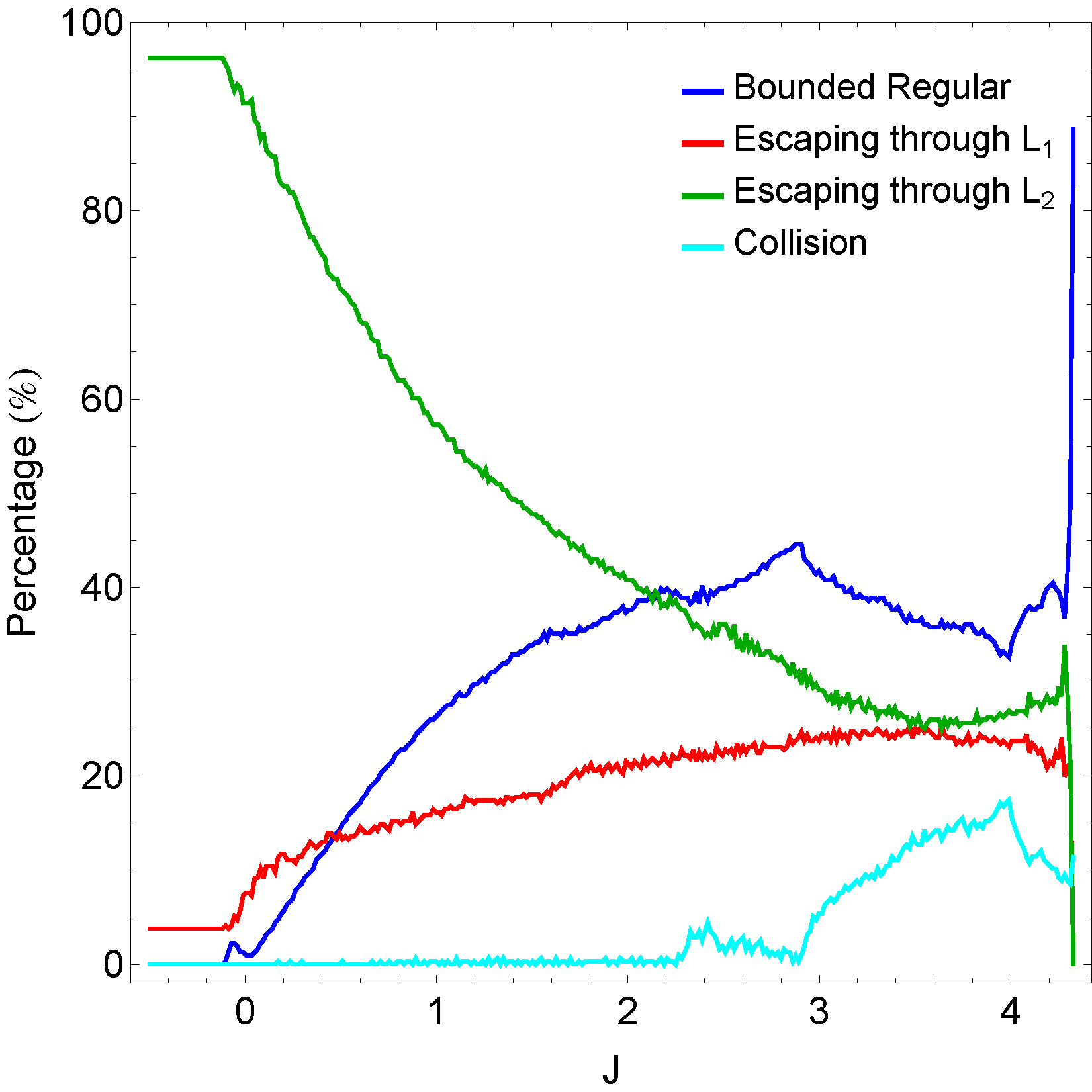}}
\caption{Evolution of the percentages of all types of orbits with initial conditions on the $(x,J)$ plane, as a function of the Jacobi constant $J$. (For the interpretation of references to colour in this figure caption and the corresponding text, the reader is referred to the electronic version of the article.)}
\label{pxC}
\end{figure}

It would be very informative to monitor how the percentages of all types of orbits evolve as a function of the value of the Jacobi integral of motion. Such a diagram is presented in Fig. \ref{pxy}. It was constructed by taking into account the information of the 16 CCDs previously described in Fig. \ref{xy}. We see that at relatively high values of $J$ $(J > 4.2)$ about half of the configuration $(x,y)$ plane is covered by initial conditions of non-escaping regular orbits. However as the value of $J$ decreases the percentage of non-escaping orbits is also reduced and for $J < 0$ it completely vanishes. The pattern of the evolution of the rate of collision orbits is very similar. The main difference is that for $J < 0$ collision motion is still possible even though the corresponding rate is extremely low (about 0.5\%). The percentage of escaping orbits which leak out through $L_1$ initially increases, while for $J < 3$ it seems to saturate around 30\%. On the other hand, the rate of escaping orbits through $L_2$ increases, almost linearly, and at the highest energy level studied $(J = -0.5)$ they dominate covering about 80\% of the $(x,y)$ plane. Taking into consideration the above-mentioned analysis, we may reasonably conclude that in the 2D Hill problem non-escaping regular orbits is the most populated type of orbits at low energy levels, while at high enough values of the energy escaping orbits dominate. In particular we observed that the probability of an orbit to leak out through $L_2$ is much more higher than through $L_1$. Therefore we may claim that exit channel 2 is more preferable in the 2D system.

The CCDs presented in Fig. \ref{xy} can provide useful information regarding the orbital structure of the 2D system however, only for some specific values of the Jacobi integral of motion. In order to overcome this limitation we shall adopt the H\'{e}non's method \cite{H69} thus using the section $y = \dot{x} = 0$, $\dot{y} > 0$. This mean that all the initial conditions of the 2D orbits will be launched from the $x$ axis, with $x = x_0$, parallel to the $y$ axis. Therefore, only orbits with pericenters on the $x$ axis will be included, while the value of the Jacobi constant $J$ can now be used as an ordinate. Using this method we are able to monitor the evolution of the orbital structure of the 2D Hill system using a continuous spectrum of energy values, rather than a few discrete ones. The orbital structure of the $(x,J)$ plane when $J \in [-0.5,J_L)$ is presented in Fig. \ref{xCt}a, while in Fig. \ref{xCt}b the distribution of the corresponding escape and collision time of the orbits is given.

With a closer look at Fig. \ref{xCt}a we can distinguish a white vertical line located at $x = 0$. We decided to exclude the initial condition $x_0 = 0$ for all energy levels because in this case the radius $R = \sqrt{x_0^2 + y_0^2}$ is zero (remember that in the $(x,J)$ plane all orbits have $y_0 = 0$) and therefore is smaller than the defined collision radius $R_{\rm col} = 10^{-4}$. This mean that all orbits with $x_0 = 0$ start inside the assumed radius of the secondary and this has no physical meaning. It is seen that at high values of the Jacobi constant, specifically for $x < 0$, a main stability islands is present. As we proceed to lower values of $J$ however its area is constantly reduced and when $J < 0$ it disappears. Escape basins on the other hand are always present, while their area grows rapidly as we move away form the critical energy level $J_L$. Basins of escape composed of initial conditions of orbits that escape through $L_2$ are much more prominent with respect to those corresponding to escape through $L_1$. It is interesting to note that Fig. \ref{xCt}a indicates how the fractality of the several basin boundaries varies not only as a function of the Jacobi constant but also
of the spatial variable. In particular, we can observe that the fractality of the basin boundaries, which is of course related to the unpredictability, migrates from the upper right side of the $(x,J)$ plane for relatively high values of $J$ (or in other words low values of the total orbital energy) to the lower left side of the same plane for low values of the Jacobi constant. According to Fig. \ref{xCt}b near the vicinity of those fractal basin boundaries the highest values of the escape time of the orbits have been reported.

In Fig. \ref{pxC} we illustrate how the percentages of all types of orbits with initial conditions on the $(x,J)$ plane of Fig. \ref{xCt} evolve as a function of the value of the Jacobi integral of motion. We see that the evolution of the rates of all types of orbits are more or less similar to that observed earlier in Fig. \ref{pxy}. The main differences between the two plots, which correspond to initial conditions of orbits in different types of planes, that occur for relatively high energy levels $(J < 0)$ are the following: (i) the rate of escaping orbits through $L_1$ in the $(x,J)$ plane seems to saturate at a lower value (at about 5\%) than in the $(x,y)$ plane which was equal to 20\%, (ii) the percentage of escaping orbits through $L_2$ is more dominant in the $(x,J)$ plane as it occupies more than 95\% of the same plane, while in the configuration $(x,y)$ plane it was found to be lower than 80\%.

\begin{figure*}[!t]
\centering
\resizebox{\hsize}{!}{\includegraphics{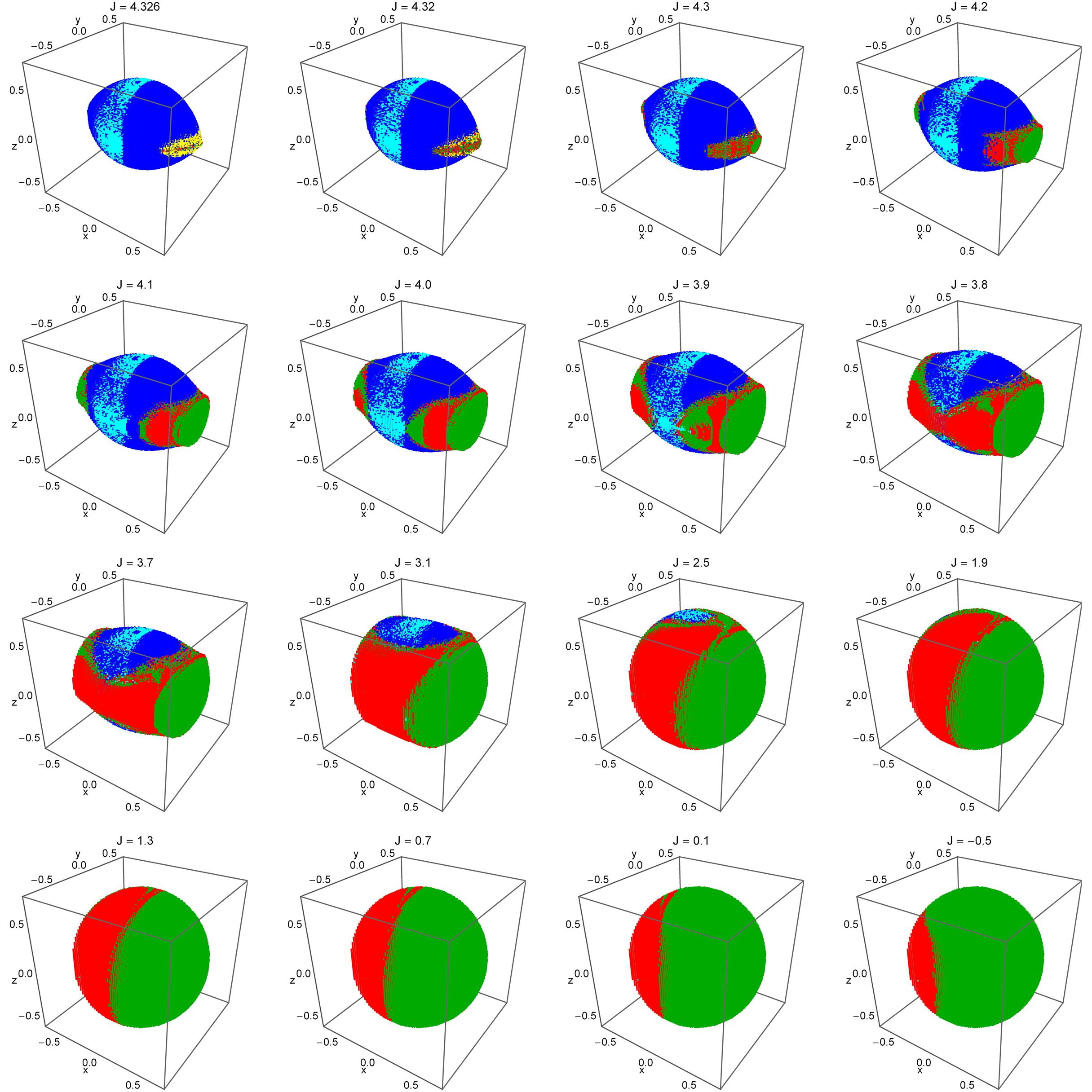}}
\caption{Evolution of the orbital structure of the configuration $(x,y,z)$ space with decreasing value of the Jacobi integral of motion. The color code is the same as in Fig. \ref{xy}. (For the interpretation of references to colour in this figure caption and the corresponding text, the reader is referred to the electronic version of the article.)}
\label{3d}
\end{figure*}

\begin{figure*}[!t]
\centering
\resizebox{\hsize}{!}{\includegraphics{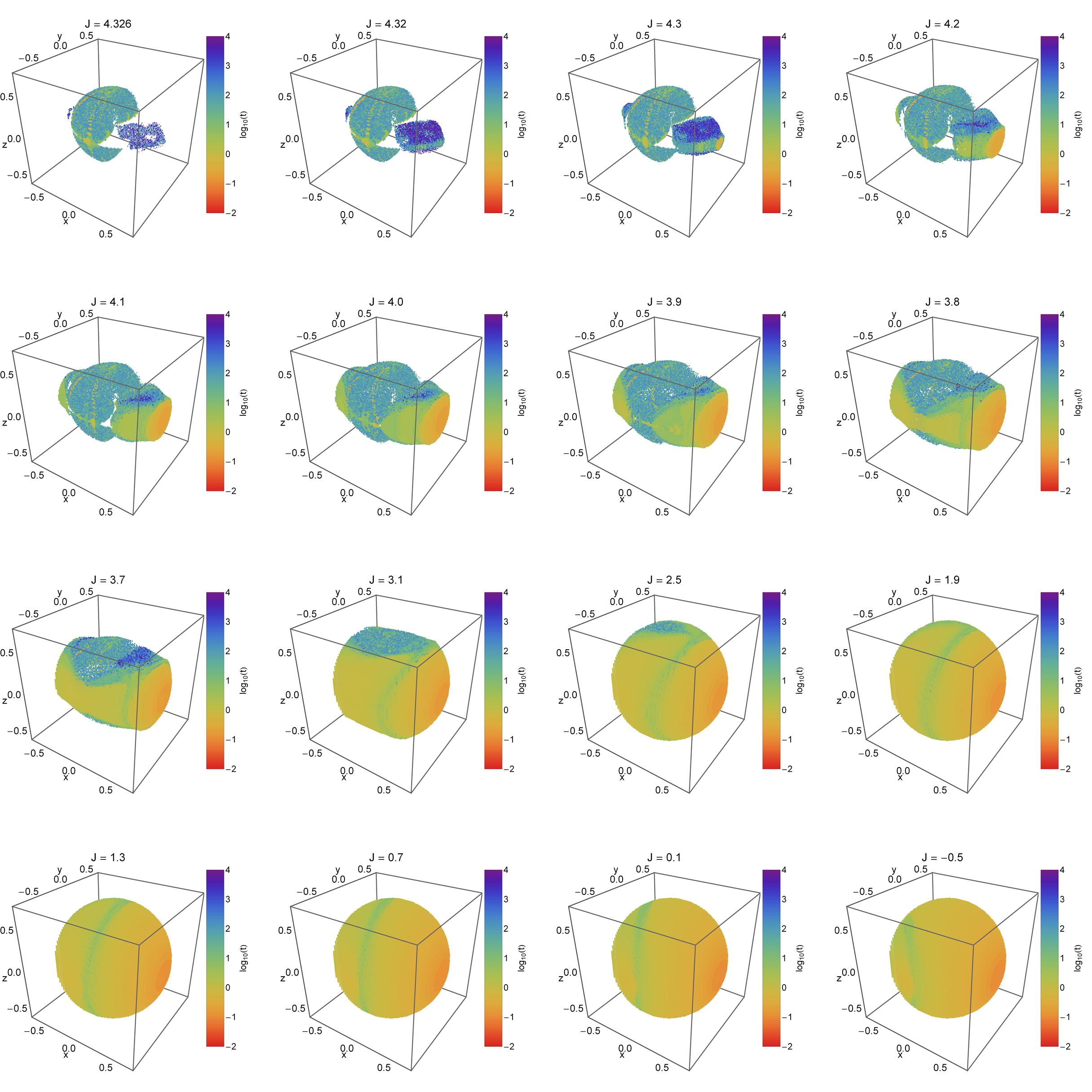}}
\caption{Distribution of the corresponding escape and the collision time of the orbits with initial conditions in the configuration $(x,y,z)$ space for the values of Jacobi integral of motion presented in Fig. \ref{3d}. The bluer the color, the larger the escape or the collision time. Initial conditions of both non-escaping regular orbits and trapped chaotic orbits are shown in transparent white. (For the interpretation of references to colour in this figure caption and the corresponding text, the reader is referred to the electronic version of the article.)}
\label{3dt}
\end{figure*}

\begin{figure*}[!t]
\centering
\resizebox{\hsize}{!}{\includegraphics{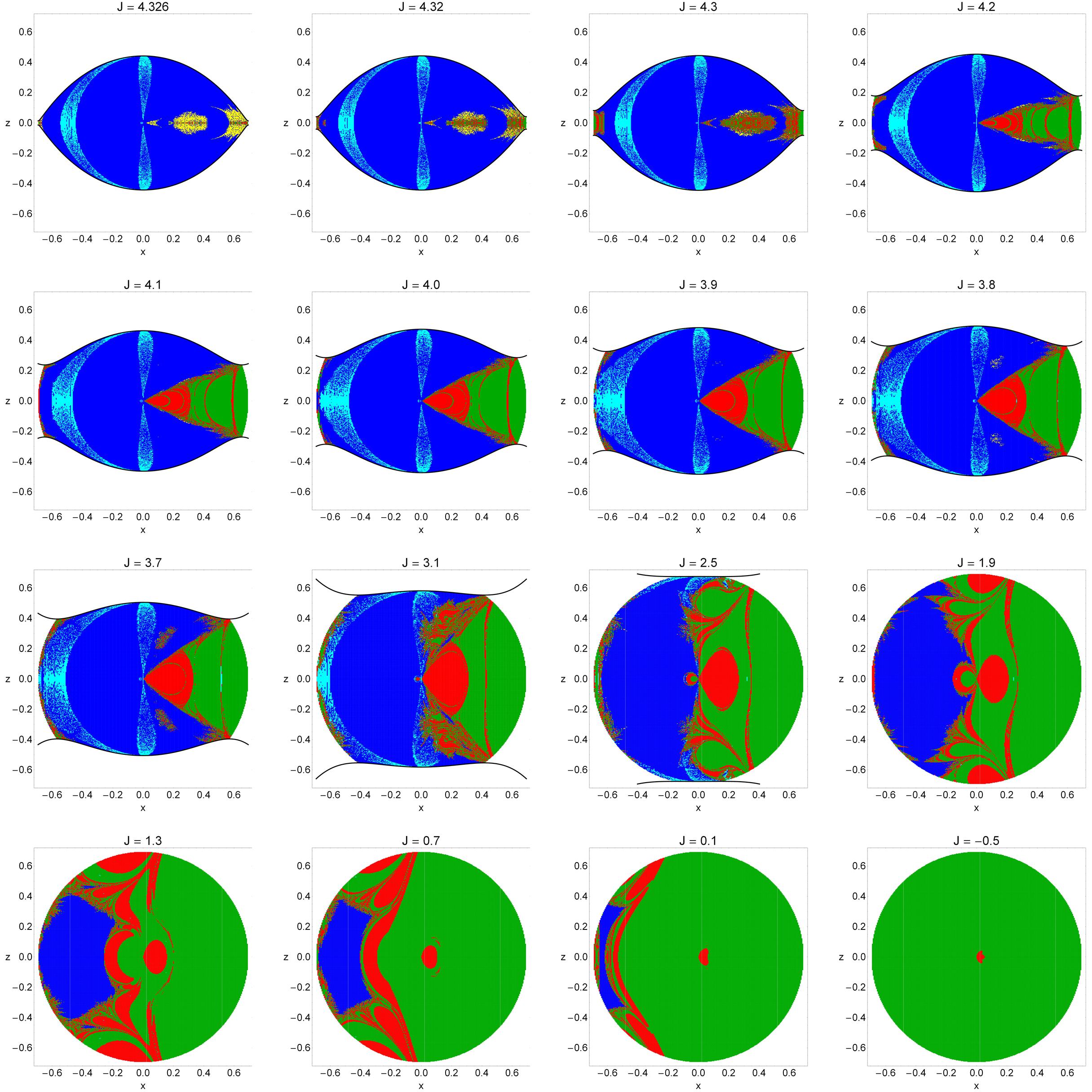}}
\caption{Evolution of the orbital structure of the $(x,z)$ space with decreasing value of the Jacobi integral of motion. The color code is the same as in Fig. \ref{xy}. (For the interpretation of references to colour in this figure caption and the corresponding text, the reader is referred to the electronic version of the article.)}
\label{xz}
\end{figure*}

\begin{figure*}[!t]
\centering
\resizebox{\hsize}{!}{\includegraphics{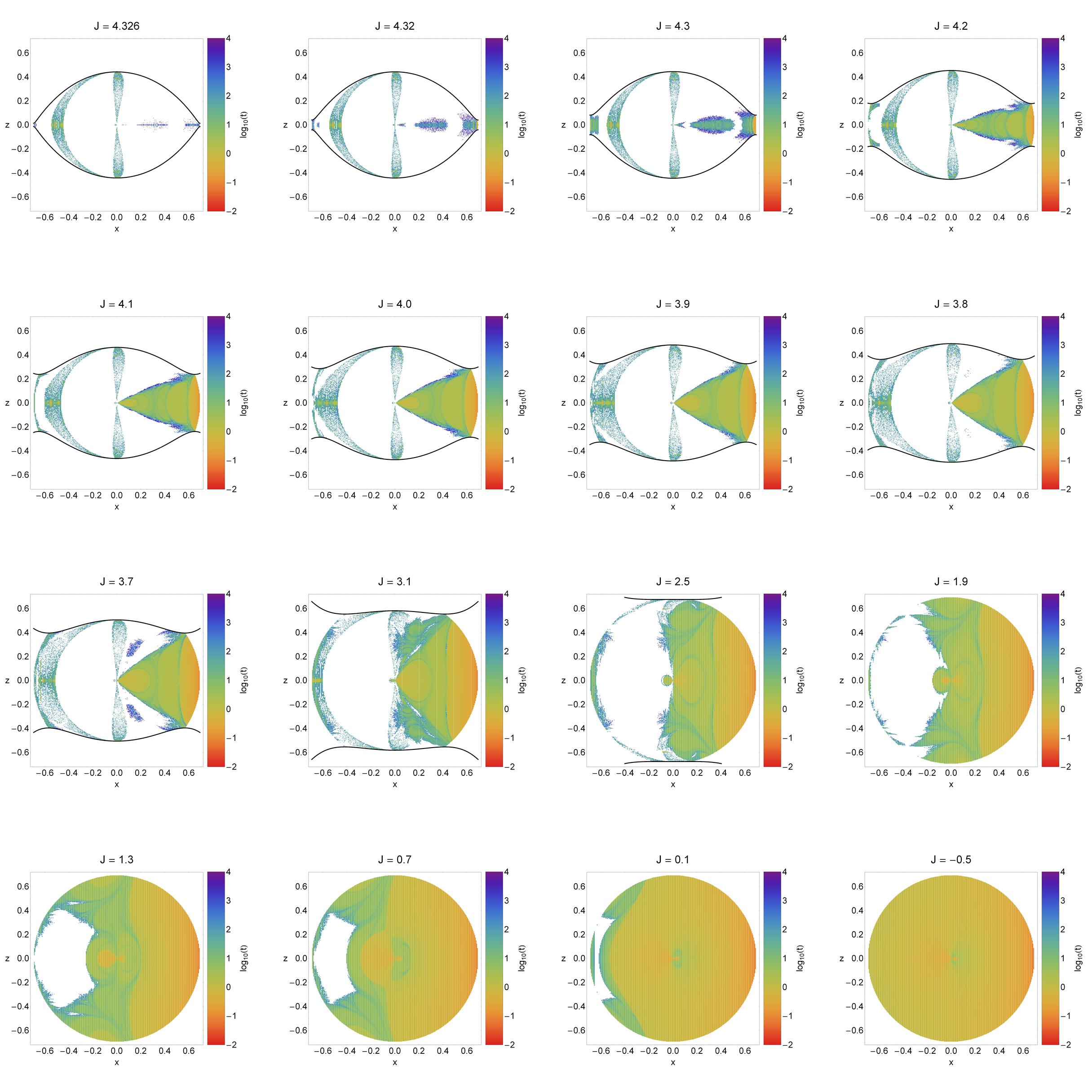}}
\caption{Distribution of the corresponding escape and the collision time of the orbits with initial conditions on the $(x,z)$ space for the values of Jacobi integral of motion presented in Fig. \ref{xz}. (For the interpretation of references to colour in this figure caption and the corresponding text, the reader is referred to the electronic version of the article.)}
\label{xzt}
\end{figure*}

\begin{figure}[!t]
\centering
\resizebox{\hsize}{!}{\includegraphics{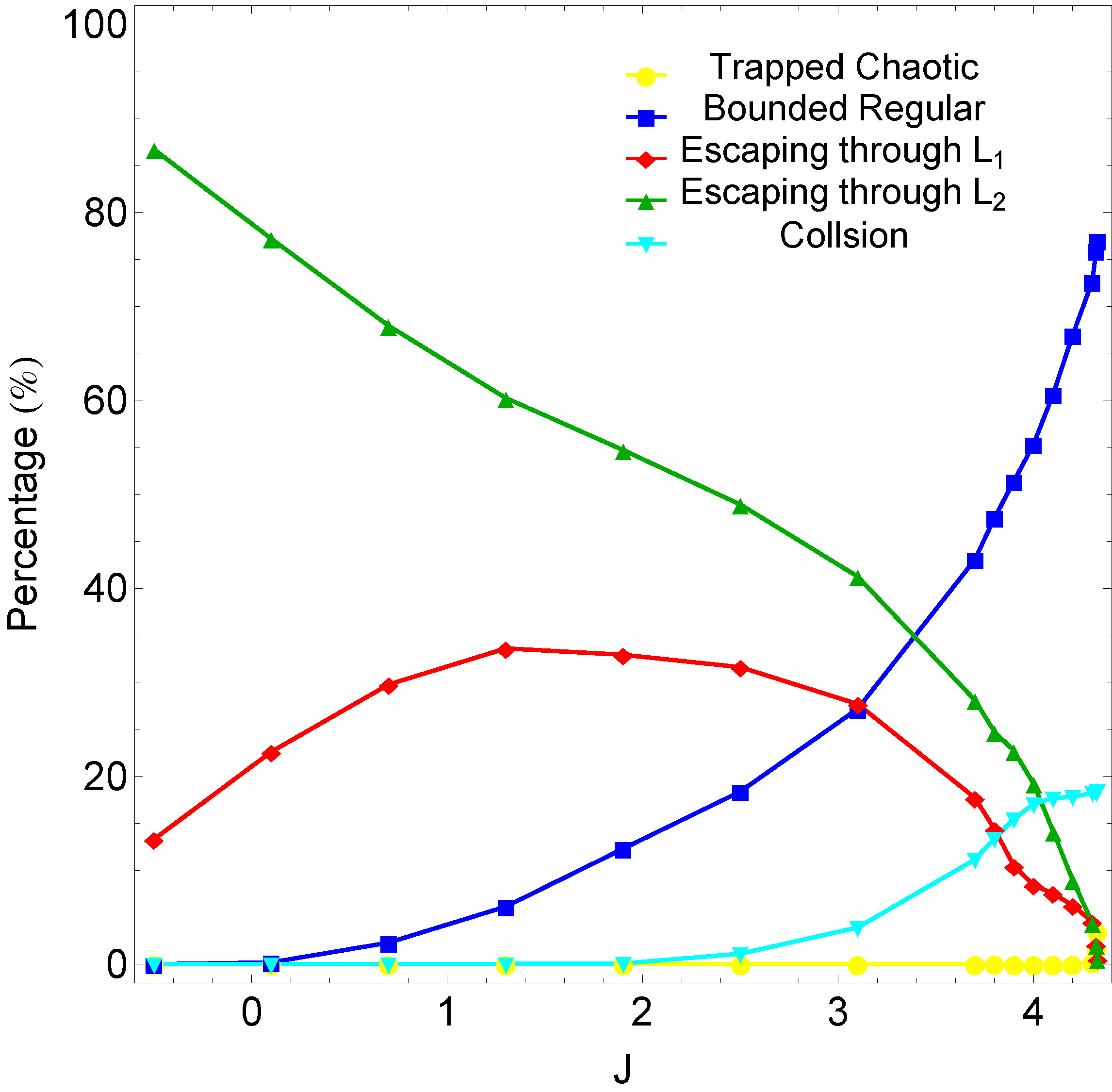}}
\caption{Evolution of the percentages of all types of orbits with initial conditions in the configuration $(x,y,z)$ space, as a function of the Jacobi constant $J$. (For the interpretation of references to colour in this figure caption and the corresponding text, the reader is referred to the electronic version of the article.)}
\label{p3d}
\end{figure}

\begin{figure*}[!t]
\centering
\resizebox{\hsize}{!}{\includegraphics{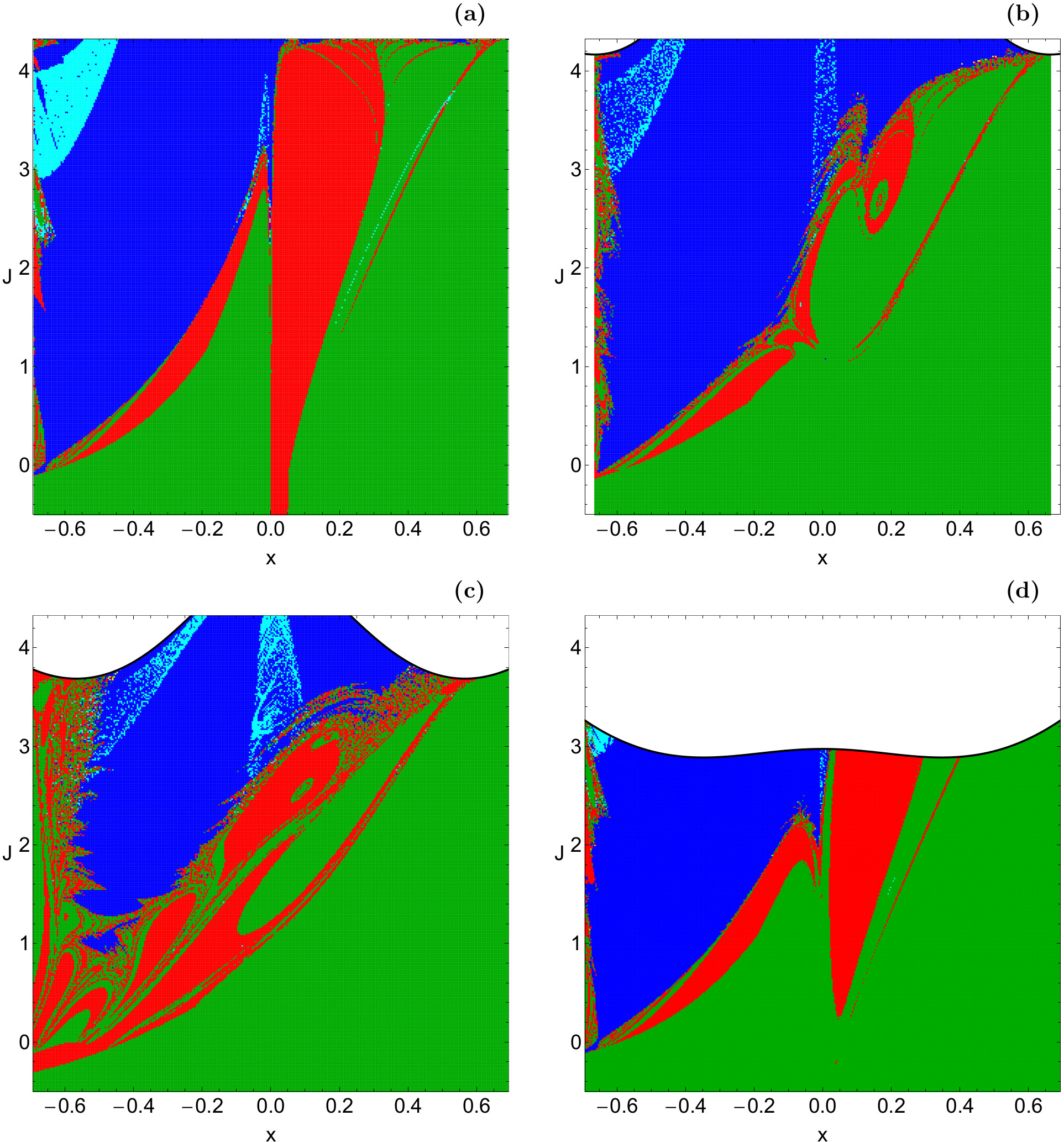}}
\caption{The orbital structure of the $(x,J)$ plane when (a-upper left): $z_0 = 0.012$, (b-upper right): $z_0 = 0.2$, (c-lower left): $z_0 = 0.4$, and (d-lower right): $z_0 = 0.6$. The color code is the same as in Fig. \ref{xy}. (For the interpretation of references to colour in this figure caption and the corresponding text, the reader is referred to the electronic version of the article.)}
\label{xC}
\end{figure*}

\begin{figure*}[!t]
\centering
\resizebox{\hsize}{!}{\includegraphics{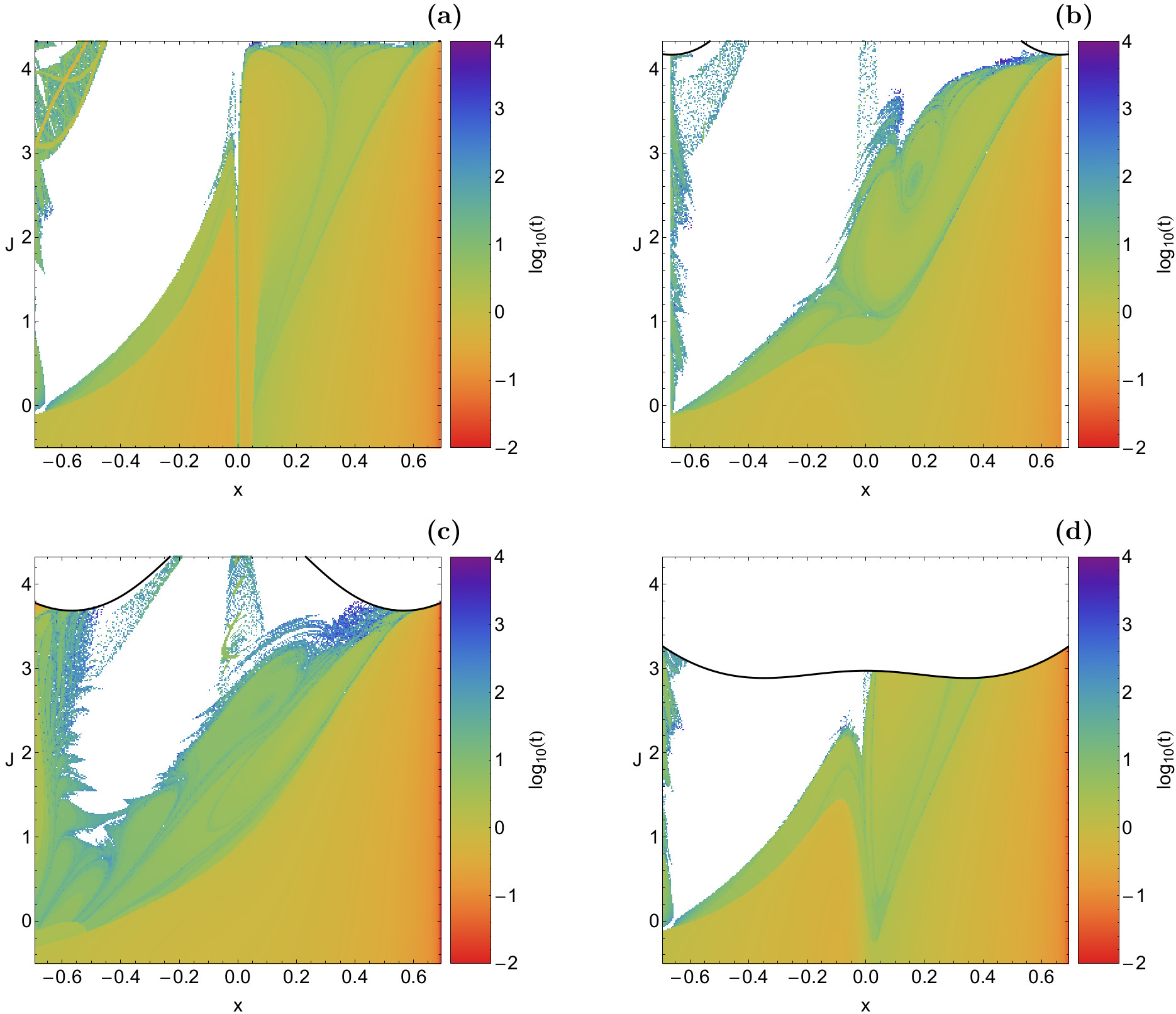}}
\caption{Distribution of the corresponding escape and the collision time of the orbits with initial conditions on the $(x,J)$ plane for the cases presented in Fig. \ref{xC}. (For the interpretation of references to colour in this figure caption and the corresponding text, the reader is referred to the electronic version of the article.)}
\label{xt}
\end{figure*}

\begin{figure*}[!t]
\centering
\resizebox{\hsize}{!}{\includegraphics{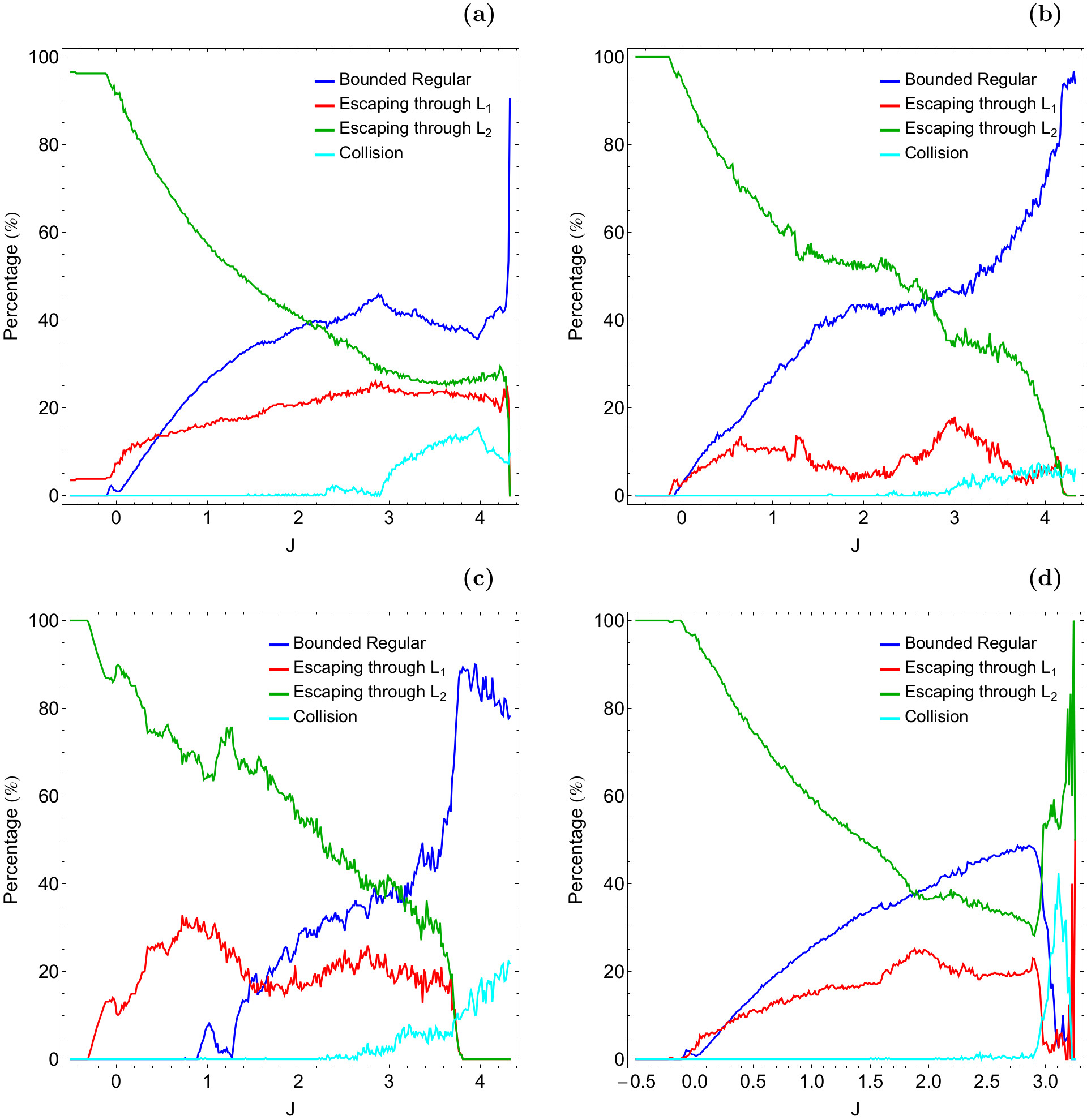}}
\caption{Evolution of the percentages of all types of orbits with initial conditions on the $(x,J)$ plane when (a-upper left): $z_0 = 0.012$, (b-upper right): $z_0 = 0.2$, (c-lower left): $z_0 = 0.4$, and (d-lower right): $z_0 = 0.6$. (For the interpretation of references to colour in this figure caption and the corresponding text, the reader is referred to the electronic version of the article.)}
\label{pxc3d}
\end{figure*}

\subsection{Results for the 3D system}
\label{ss2}

Our numerical investigation continues with the full 3D system. Fig. \ref{3d} presents a collection of CCDs using three dimensional distributions of orbits with initial condition inside the $(x,y,z)$ space for the same energy levels of Fig. \ref{xy}. We observe that in this case the grids composed of the initial conditions of the orbits are in fact three dimensional solids which means that only their outer shell is visible. Three dimensional distributions of initial conditions of orbits have also been recently used in order to reveal the orbital structure of the 3D Earth-Moon system \cite{Z16a}. The corresponding distributions of the escape and collision time of the 3D orbits are given in Fig. \ref{3dt}, where for the initial conditions of the non-escaping regular orbits as well as the trapped chaotic ones we have used transparent white color in order to be able to inspect, in a way, the interior of the solids.

If we want to examine in more detail the interior region of the 3D grids we have to use a tomographic-style approach as in \cite{Z14a}. According to this approach we take the cuts on the $(x,y)$ plane for several levels of the $z$ coordinate. This method however is very costly as for every 3D grid we need several 2D cuts on the $(x,y)$ plane. For saving space we decided to present for every energy level a cut on the $(x,z)$ plane. Therefore all 3D orbits with initial conditions on the $(x,z)$ plane have $y_0 = \dot{x_0} = \dot{z_0} = 0$, while the initial value of $\dot{y}$ is always obtained from the Jacobi integral of motion (\ref{ham}). There is also one more reason justifying our choice regarding the type of the 3D cuts. The $z$ coordinate is directly related with the 3D motion of orbits, therefore we decided to use a continuous spectrum of $z$ values rather than a few cuts for specific values of $z_0$ as in \cite{Z14a}. Furthermore, the CCDs of the 2D system presented earlier in Fig. \ref{xy} are in fact the $z = 0$ cuts of the corresponding 3D solids. At this point, we would like to stress out that as far as we know this is the first time that 3D distribution of initial conditions of orbits are systematically classified in the Hill problem.

The evolution of the orbital structure of the $(x,z)$ space, as the value of the Jacobi integral decreases, is revealed through the rich collection of the CCDs presented in Fig. \ref{xz}, while the corresponding distribution of the escape and collision time of the orbits is given in Fig. \ref{xzt}. The outermost black solid line denotes the zero velocity curve which is defined as $2W(x,y=0,z) = J$. It is seen that the orbital structure is very different with respect to that observed in the 2D system. For high values of the Jacobi constant non-escaping regular motion dominates, while initial conditions of escaping orbits are mainly confined to the right side of the $(x,z)$ planes. A very interesting phenomenon is the presence of trapped chaos. Trapped chaotic orbits have also been observed in other types of 3D open Hamiltonian systems. For instance, in \cite{Z15a} we detected a considerable amount of trapped chaotic orbits when we investigated the orbital properties of an open tidally limited star cluster, using a three dimensional dynamical model. On the other hand, in the 3D Earth-Moon system which was explored in \cite{Z16a} we did not find any numerical evidence of trapped chaos. The most important changes that occur on the $(x,z)$ plane as the value of the Jacobi constant decreases are the following:
\begin{itemize}
  \item The area composed of initial conditions corresponding to non-escaping regular orbits is constantly reduced. It is seen that for $J < 2.5$ bounded basins are present only at the left side of the $(x,z)$ plane, while for negative values of the Jacobi integral of motion they completely disappear.
  \item Collision basins exist mainly inside the vast regular regions thus creating a complicated mixture of initial conditions of both collision and non-escaping regular orbits. We found that for $J < 1.9$ there is no indication of collision motion whatsoever.
  \item Initial conditions corresponding to trapped chaotic orbits are visible only for relatively high values of the Jacobi constant, very close to the critical energy value $J_L$. Our computations suggest that for $J < 4.2$ the amount of trapped chaotic orbits is extremely low and the corresponding initial conditions appear as lonely points randomly scattered in the $(x,z)$ plane and therefore they are not visible.
  \item As we move away from the critical value of the energy to higher levels of energy the basins of escape start to expand. Initially they are mainly located to the right side of the $(x,z)$ plane however as the value of $J$ decreases they migrate also to the left side of the same plane. Finally at extremely low levels of $J$ and particularly for negative values basins of escape cover the entire $(x,z)$ plane.
  \item Our numerical calculations strongly indicate that the two exit channels, even though they are symmetrical, they are not equiprobable. Indeed it was found that the test particles have the tendency to leak out mainly from the exit channel near $L_2$. Similar behavior was also observed in the 2D system. It is interesting to note that for extremely low values of the Jacobi integral of motion $(J < 0)$ almost the entire $(x,z)$ plane is dominated by initial conditions of orbits which escape through channel 2. However, a small basin of escape corresponding to channel 1 is still present near the origin.
\end{itemize}

Exploiting the numerical integration of the three dimensional distributions of initial conditions of orbits shown in Fig. \ref{3d} we managed to demonstrate in the diagram of Fig. \ref{p3d} the evolution of the percentages of all types of orbits as a function of the value of the Jacobi integral of motion. It is seen that at high enough values of $J$, just above the energy of escape, about 80\% of the configuration $(x,y,z)$ space is covered by initial conditions of non-escaping regular orbits. As the value of $J$ decreases however the rate of regular orbits also decreases until it completely disappears for $J < 0$. The evolution of the rate of collision orbits is quite similar as it starts at about 18\% for $J > 4.3$ and then it drops tending asymptotically to zero for relatively high values of the energy. Here it should be pointed out that additional numerical calculations (not shown here) suggest that collision motion remains mathematically possible, even for extremely low values of the Jacobi constant, even though the corresponding percentage is lower than 0.01\%. The percentage of escaping orbits through $L_1$ initially increases until $J = 1.3$ where it exhibits its maximum value at around 33.5\%, while for lower values of $J$ the tendency is reversed. On the other hand, the rate of orbits that leak out through $L_2$ is constantly increases. At the highest energy level studied, that is for $J = -0.5$, escaping orbits through channel 2 cover about 86\% of the configuration space, while at the same time orbits that escape through exit channel 1 occupy only 13\% of the same space. Thus is becomes more that evident that the escape process in the 3D Hill system is a very complicated phenomenon.

The CCDs presented in both Figs. \ref{3d} and \ref{xz} reveal the orbital properties of the 3D system however for a fixed value of the Jacobi integral of motion. In order to obtain a more complete view of the orbital structure of the 3D Hill system, we shall try to gather information of a continuous spectrum of energy values. To achieve this we will adopt the approach used previously for the 2D system. In particular, for specific values of $z_0$ we define dense uniform grids of initial conditions on the $(x,J)$ plane with $y_0 = \dot{x_0} = \dot{z_0} = 0$, while in all cases the initial value of $\dot{y}$ is obtained from the energy integral (\ref{ham}). Fig. \ref{xC}a shows the case when $z_0 = 0.012$. We observe that the orbital structure is quite similar to that shown in Fig. \ref{xCt}a. This however is anticipated because the particular value of $z_0$ is relatively low and therefore the motion of the orbits takes place very close to the two-dimensional $(x,y)$ plane. When $z_0 = 0.2$ the pattern changes drastically. The collision basin located at the upper left part of the $(x,J)$ plane seems to weaken, since initial conditions of non-escaping regular orbits emerge inside it. Furthermore it is seen that basins composed of bounded regular orbits expand also at the right side of the plane, while the extent of the basins of initial conditions of orbits that escape through $L_1$ is reduced. The outermost black solid line is the limiting curve which is defined as
\begin{equation}
f_1(x,J;z_0) = 2W(x,y = 0,z = z_0) = J.
\label{zvc1}
\end{equation}
In panel (c) where $z_0 = 0.4$ it is interesting to note the rich fractal boundaries between the several basins of escaping and non-escaping regular orbits. On the contrary in panel (d) where $z_0 = 0.6$ the degree of fractality is considerable lower since the basin boundaries are much more smooth. In addition, the basins of collision orbits seem to be confined only to the upper left side of the $(x,J)$ plane. It should be clarified that in this case (the 3D system) the vertical white line corresponding to initial conditions of orbits inside the collision radius of the secondary is not present. This is true because all four values of $z_0$ are such $(z_0 > 0.01)$ so the distance $R = \sqrt{x_0^2 + y_0^2 + z_0^2}$ is always larger than the collision radius, even when $x_0 = y_0 = 0$. Looking the four panels of Fig. \ref{xC}(a-d) it becomes evident that the energetically permissible area on the $(x,J)$ plane is reduced as the initial value of the $z$ coordinate increases. The distribution of the corresponding escape and collision time of the orbits with initial conditions on the $(x,J)$ plane is presented in Fig. \ref{xt}(a-d). We see in panel (c) that the escape time of the orbits with initial conditions in the vicinity of the fractal basin boundaries are extremely high corresponding to more than 1000 time units. On the other hand, all orbits with initial conditions very close to the equilibrium point $L_2$ (outer right side of the plots) escape almost immediately within the first couple of time steps of the numerical integration.

We close our numerical investigation by presenting in Fig. \ref{pxc3d}(a-d) the evolution of the percentages of all types of orbits with initial conditions on the $(x,J)$ plane for the four cases discussed in Fig. \ref{xC}(a-d). One may observe that in all cases escaping orbits completely dominate for extremely low values of the Jacobi integral of motion, or in other words for high values of the total orbital energy. For high values of $J$ on the other hand, non-escaping regular orbits is the most populated type of orbits in the first three cases, except the case where $z_0 = 0.6$. In this case escaping orbits dominate but this is actually a numerical artifact created by the strange geometry of the limiting curve (see Fig. \ref{xC}d). Here we would like to clarify that the evolution of the percentages of trapped chaotic orbits was not included in all four cases because the corresponding percentage was measured to be always extremely low (lower than 0.1\%). Taking into account all the results given in Fig. \ref{pxc3d}(a-d) we may reasonably conclude that the behavior according to which for high enough values of the energy the probability an orbit to escape through exit channel 2 is considerably much higher than trough $L_1$, that initially observed in the 2D system, is still valid also in the 3D system.

\begin{figure}[!t]
\centering
\resizebox{\hsize}{!}{\includegraphics{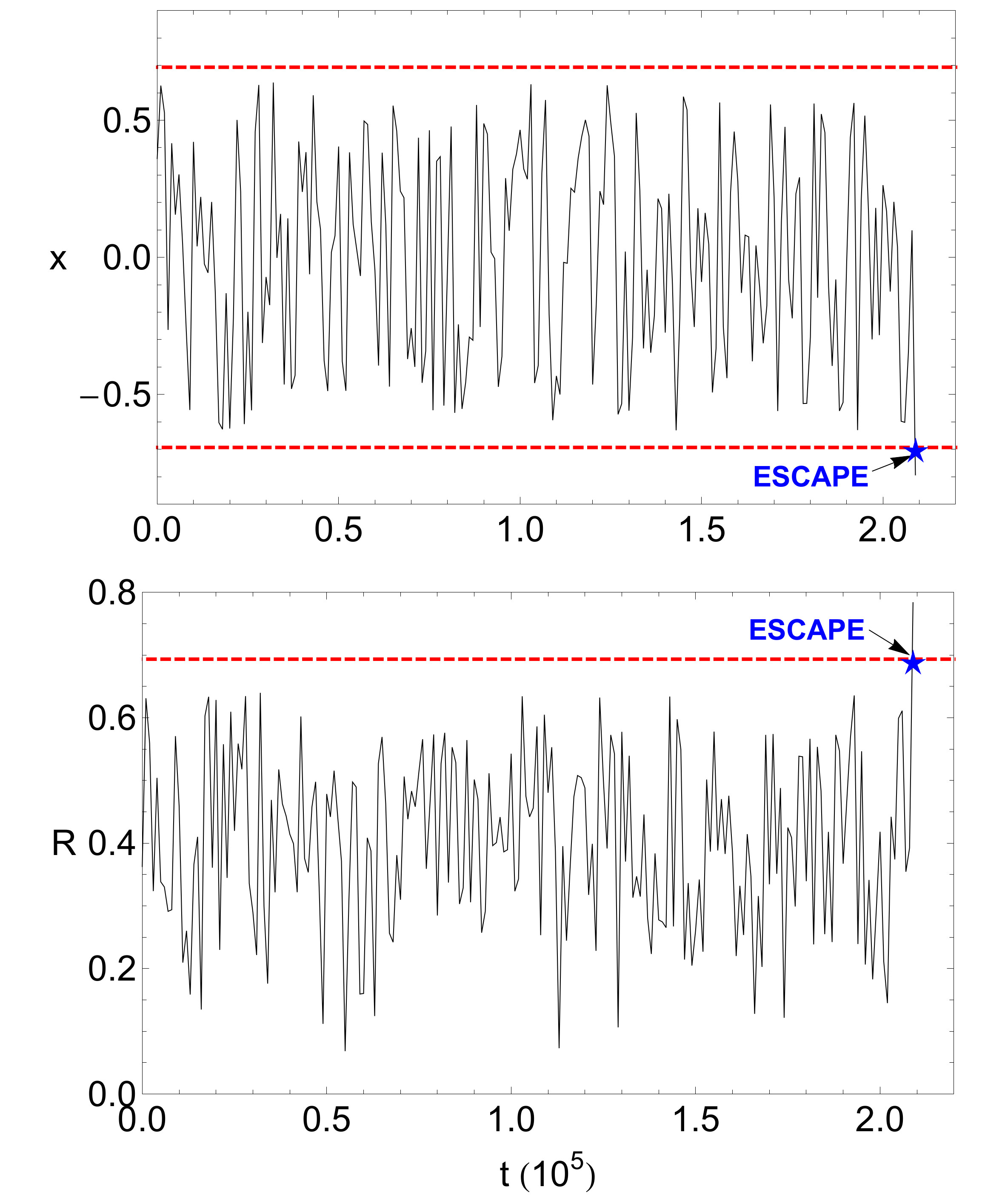}}
\caption{Time-evolution of the $x$ coordinate and the distance $R = \sqrt{x^2 + y^2 + z^2}$ from the center of a trapped chaotic orbit, when $J = 4.326$. The red vertical dashed lines denote the position of the two equilibrium points which delimit the escape of the orbits. (For the interpretation of references to colour in this figure caption and the corresponding text, the reader is referred to the electronic version of the article.)}
\label{trape}
\end{figure}

\begin{figure*}[!t]
\centering
\resizebox{\hsize}{!}{\includegraphics{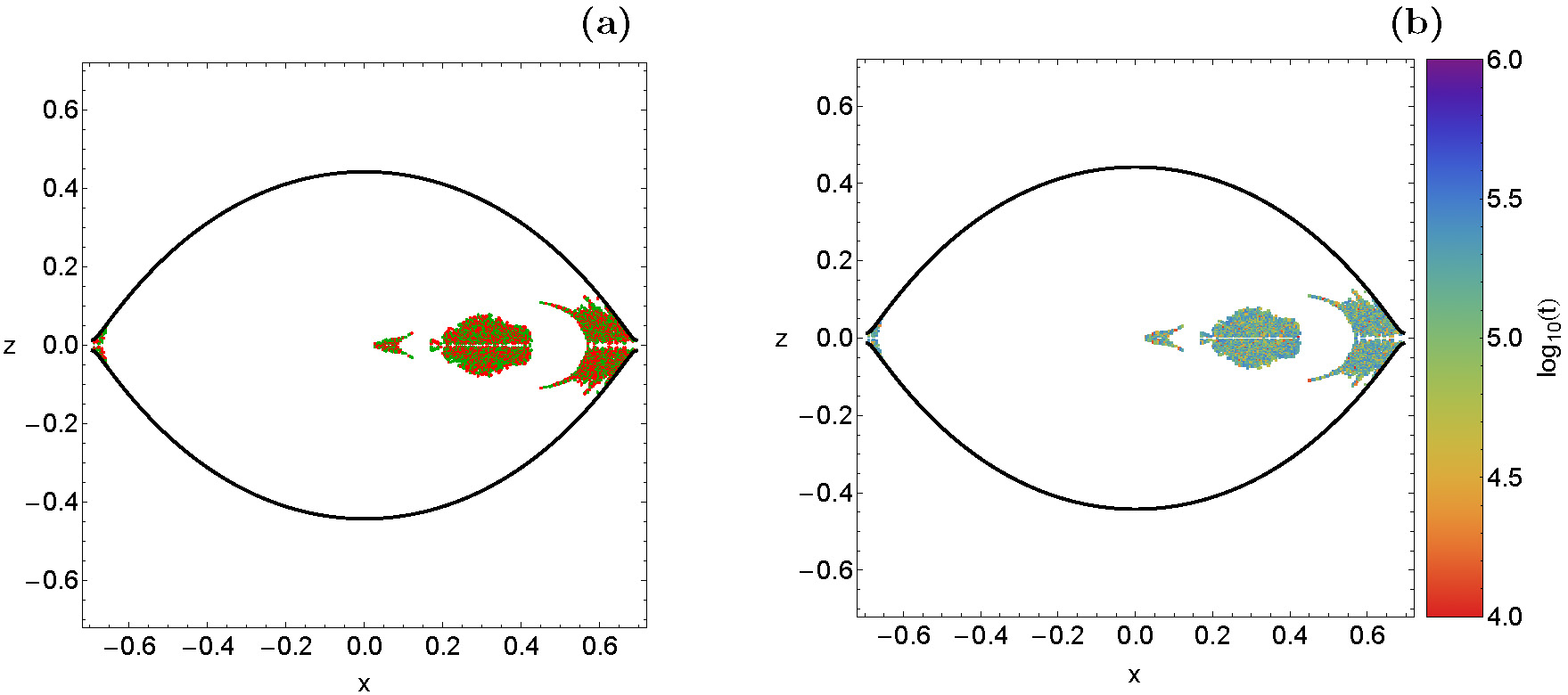}}
\caption{(a-left): Orbital structure of the $(x,z)$ plane for $J = 4.326$, when the integration time of the 3D orbits is equal to $10^6$ time units. The color code is the same as in Fig. \ref{xy}. (b-right): The distribution of the corresponding escape time of the orbits. (For the interpretation of references to colour in this figure caption and the corresponding text, the reader is referred to the electronic version of the article.)}
\label{trap}
\end{figure*}

\section{Trapped chaos}
\label{trapc}

For 3-dof Hamiltonian systems the general situation for moderate perturbation is the following: There are many surviving KAM tori which fill a part of the energy shell whose measure is larger than zero and even can be large. These invariant surfaces are 3 dimensional and therefore in the 5 dimensional energy shell they do not divide anything. I.e. the complement of all these tori is connected. On the other hand, there is an infinity of resonance zones which contain unstable behaviour. In
principle, all these resonance zones (fine chaos strips) are connected and form a single web of resonances (Arnold web). For perturbation different from zero, i.e. for systems which are not integrable, the measure of this web of resonance zones also is larger than zero. It is a fractal web of an infinity of small layers. Accordingly, at very large times an orbit starting in any initial point of this web can reach any other point of the web, even when it usually may take an extremely long time. The motion in the web looks similar to diffusive motion and for moderate perturbation the diffusion constant is very small. The surviving 3 dimensional tori form a kind of fractal of obstacles for the general motion in the space between the tori.

Now if we imagine the situation where there is also a small escape channel for escape. This escape channel is in contact with the web of chaotic resonance zones. Therefore, any orbit starting in any of the resonance zones will finally escape after a possibly extremely large time. However, when we observe for a finite time only (this may be a large time, even large compared to the life time of the universe) then we only observe the diffusive like chaotic motion inside of the web and for most orbits we do not see the final escape. For all practical purposes we have the impression having observed trapped chaos.

In the CCDs presented in Fig. \ref{xz} we observed that for energy levels just above the critical energy of escape $J_L$ there is an amount of trapped chaotic orbits. The phenomenon of trapped chaos has been also observed in other types of Hamiltonian systems (e.g., \cite{JZ16,Z15a}). According to classical chaos theory however, chaotic orbits do not admit a third integral of motion. Therefore for time $t \to \infty$ they will fill all the available phase space inside the zero velocity surface, unless of course they escape. In other words, all trapped chaotic orbits will eventually leak out passing through one of the two escape channels, given enough time of numerical integration. In Fig. \ref{trape} we present a characteristic example of a trapped chaotic orbit with initial conditions: $x_0 = 0.36005407$, $y_0 = 0$, $z_0 = 0.04320648$, $\dot{x_0} = \dot{z_0} = 0$, while the initial value of $\dot{y} > 0$ is obtained from the energy integral (\ref{ham}), when $J = 4.326$. The total time of the numerical integration this time was equal to $10^6$ time units. It is seen that the orbit does escape through exit channel 1, after about 208934 time units.

To check whether all trapped chaotic orbits, observed in the corresponding CCD (when $J = 4.326$), eventually escape or not, we numerically integrated the initial conditions of the 3D trapped chaotic orbits for $10^6$ time units. Our results are illustrated in Fig. \ref{trap}a where one can observe that all these initial conditions, that were initially classified as trapped chaotic ones for $10^4$ time units, they have escaped creating a highly fractal escape mixture. The distribution of the corresponding escape time of the 3D orbits is given in panel (b) of Fig. \ref{trap}, where this time the values on the color bar vary from 4 to 6 (remember that the scale of the accompanying color bar is in logarithmic scale). The same procedure was carried out for all the other sets of trapped chaotic orbits reported in the CCDs for lower values of the Jacobi constant. We found that all of these orbits eventually escape through one of the two channels however after extremely long time of numerical integration ($t_{\rm esc} \in (10^4, 10^6)$). Therefore, we may reasonably conclude that in the 3D Hill system long-lived (or temporarily trapped) chaotic orbits exist for values of energy above but very close to the energy of escape.

In 2-dof systems we have 3 dimensional energy shells and the 2 dimensional KAM tori separate these energy shells in disjoint inner and an outer parts. In general, there are chaotic layers also inside of the KAM tori. But these layers are disconnected from the outside. Therefore, chaos in these inside layers can never reach the exit and consequently it can never escape. Thus it is truly trapped chaos for infinite time in the strict mathematical sense. Here it should be emphasized that none of the examined initial conditions of the 3D trapped chaotic orbits in the CCDs has $z_0 = 0$. This fact supports our findings of the 2D system, where there was no numerical indication of 2D (temporarily) trapped chaotic motion whatsoever (e.g., \cite{ABSF03}).

\section{Discussion and conclusions}
\label{disc}

The aim of this paper was to reveal the overall orbital structure of the classical Hill problem. By performing a thorough and extensive numerical investigation using large sets of initial conditions, in both the two-dimensional and the three-dimensional configuration space, we managed to classify orbits into four categories: (i) non-escaping regular, (ii) trapped chaotic, (iii) escaping, and (iv) collision. We also related the several basins of escape and collision with the corresponding escape and collision time of the orbits. As far as we know, this is the first detailed and systematic numerical analysis regarding orbit classification in the classical Hill problem (especially in the 3D configuration space) and therefore this is exactly the contribution as well as the novelty of our study.

A Quad-Core i7 2.4 GHz PC was used for the numerical integration of the sets of the initial conditions of the orbits. For the completion of every set of initial conditions of orbits we needed between about 2 minutes and 4 days of CPU time, depending of course on the collision and escape time of the orbits in each case. Here it should be explained that, for saving time, when an orbit escaped or collided the numerical integration was stopped and proceeded to next available initial condition.

In this paper we provide quantitative information regarding the orbital structure of the classical planar (2D) and spatial (3D Hill problem. The main novel results of our orbit classification can be summarized as follows:
\begin{enumerate}
  \item At high enough values of the Jacobi integral of motion, or in other words at low values of the total orbital energy non-escaping regular orbits as well as collision orbits dominate in both the 2D and the 3D system, while initial conditions of escaping orbits were found mainly near the equilibrium points.
  \item Escaping orbits are the most populated type of orbits at relatively high values of the total orbital energy, where basins of escape were found to occupy almost the entire configuration space.
  \item As the value of the Jacobi constant decreases the percentages of both non-escaping regular orbits and collision orbits are reduced. Our computations suggest that for negative values of $J$ there is no indication of bounded motion whatsoever, while on the other hand collision motion is still possible.
  \item In the 3D system, and especially for values of energy above yet very close to the critical energy of escape, we identified a portion of trapped chaotic orbits. Additional numerical calculations revealed that all these orbits eventually do escape, while having huge escape periods (more than $10^4$ time units). It should be emphasized that the phenomenon of (temporarily) trapped chaos was not observed in the 2D system.
  \item In both the 2D and the 3D systems we observed that the two symmetrical escape channels located near the equilibrium points are not equiprobable\footnote{The two saddle points $L_1$ and $L_2$ lie along the $x$ axis and the zero velocity surfaces are symmetric with respect to the three axes. It is easy to prove that if $(x(t), y(t), z(t))$ is a solution of the Hill problem, $(-x(t), -y(t), -z(t))$ is also a solution, of course with appropriately chosen momenta (because of the presence of the Coriolis force). Therefore, for every orbit escaping through $L_1$ there is a symmetric-related orbit escaping through $L_2$, because the dynamics of the system must be symmetric. On this basis, the observed preference for escape through exit channel 2 is due to the particular choice of the initial conditions. Similar choices of initial conditions of orbits can be found in several earlier related papers (e.g., \cite{N04,N05,dAT14,Z15f,Z15g,Z16a})}. In fact, for low values of the Jacobi constant the vast majority of orbits choose to escape through exit channel 2, while the percentage of exit channel 1 remains very low.
  \end{enumerate}

For very low values of the mass ratio $\mu$ the restricted three-body problem is well approximated by the the Hill problem. It should be emphasized that there are numerous previous works on the field of capture and escape on the RTBP (e.g., \cite{AF04,RS07}) as well as on the Hill problem (e.g., \cite{WBW04,WBW05}). Furthermore, similar systematic classification of initial conditions of orbits, with respect to the present one, has been performed in the RTBP in general (e.g., \cite{N04,N05,Z15c,Z15d,Z15e}) and also in specific planetary systems (e.g., \cite{dAT14,Z16a}) (Earth-Moon system), (e.g., \cite{Z15f}) (Saturn-Titan system), and (e.g., \cite{Z15g}) (Pluto-Charon system). Therefore it would be very interesting if we could determine the similarities and the differences between the RTBP and the Hill problems. However there are two main reasons which make this task impossible: (i) the scattering region is not the same in all these works. In some cases the scattering region is extended around both primaries, while in some other cases it is confined very close to the secondary. (ii) the criteria for distinguishing between the several types of the orbits are different. For instance, in the present Hill problem there is only one type of escaping orbits, while on the other hand in the previous cases there were two types of escape (escape toward the primary realm and escape toward the exterior region). Therefore, we believe that if we try to compare all the different results we might end to erroneous and/or misleading conclusions due to the two above-mentioned reasons.

Taking into account the novel outcomes of our detailed numerical investigation we could argue that out task has been successfully completed. We hope that the present results to be useful in the field of the orbital dynamics of the Hill problem. In the next two papers of the series we will explore the orbital structure of the Hill problem where the perturbations of the oblateness (Part II) and the radiation pressure (Part III) are present.

\section*{Acknowledgments}

The author would like to express his warmest thanks to the three anonymous referees for the careful reading of the original manuscript and for all the apt suggestions and comments which allowed us to improve both the quality as well as the clarity of the paper.

\section*{Compliance with Ethical Standards}

\begin{itemize}
  \item Funding: The author states that he has not received any research grants.
  \item Conflict of interest: The author declares that he has no conflict of interest.
\end{itemize}

\section*{Appendix: Derivation of the potential function of the classical Hill problem}
\label{apex}

The first step is to perform the coordinate transformation given in Eq. (\ref{trans0}) along with $\mu = M^3$. Then Eq. (\ref{pot}) becomes a polynomial function $\Omega = \Omega(x,y,z,M)$.

It is very easy to prove that $\Omega_{\rm lim} = \displaystyle{\lim_{M \to 0} \Omega} = 3/2$.

Next we expand $\Omega$ into a power series around $M = 0$ (Taylor series), keeping terms up to second order thus having
\begin{equation}
W_0(x,y,z,M) = \frac{3}{2} + M^2 \left(\frac{3x^2}{2} - \frac{z^2}{2} + \frac{1}{r}\right),
\label{texp}
\end{equation}
where $r = \sqrt{x^2 + y^2 + z^2}$.

Now for obtaining the potential function $W(x,y,z)$ of the classical Hill problem all we have to do is to eliminate the parameter $M$ for Eq. (\ref{texp}). This can be achieved as
\begin{equation}
W(x,y,z) = \frac{W_0 - \Omega_{\rm lim}}{M^2},
\end{equation}
thus deriving the final form of Eq. (\ref{pot2})
\begin{equation}
W(x,y,z) = \frac{3x^2}{2} - \frac{z^2}{2} + \frac{1}{r}.
\end{equation}

\end{document}